\documentclass[12pt,preprint]{aastex}

\newcommand{\lw}[1]{\smash{\lower2.ex\hbox{#1}}}
\def\kms{km~s$^{-1}$}

\def\hii{{\rm H}{\scriptsize{\rm II}}}

\def\kms{\mbox{km~s$^{-1}$}}
\def\cmc{cm$^{-3}$}
\def\cmq{cm$^{-2}$}

\def\um{$\mu$m}

\def\Msun{\mbox{$M_\odot$}}
\def\Lsun{\mbox{$L_\odot$}}
\def\Vlsr{$V_{\rm LSR}$}
\def\Vsys{$V_{\rm sys}$}

\def\Vrot{$V_{\rm rot}$}
\def\mic{\mbox{$\mu$m}}

\def\nodata{$\cdot\cdot\cdot$}

\def\Snu{$S_{\nu}$}

\def\wat{H$_2$O}
\def\amm{NH$_3$}

\def\HCO{\mbox{HCO$^+$}}

\def\mcn{\mbox{CH$_3$CN}}
\def\Xmcn{\mbox{$X_{\rm CH_3CN}$}}
\def\Nmcn{\mbox{$N_{\rm CH_3CN}$}}

\def\co{$^{12}$CO}
\def\tCO{$^{13}$CO}
\def\CeO{C$^{18}$O}
\def\NtwoH{N$_2$H$^+$}

\def\pbeam{beam$^{-1}$}

\def\mic{\mbox{$\mu$m}}
\def\sgm{\mbox{$\sigma$}}
\def\Reff{\mbox{$R_{\rm eff}$}}
\def\Req{\mbox{$R_{\rm eq}$}}
\def\dVint{\mbox{$\Delta v_{\rm int}$}}

\def\Mlte{\mbox{$M_{\rm LTE}$}}
\def\Mvir{\mbox{$M_{\rm vir}$}}
\def\Mdust{\mbox{$M_{\rm dust}$}}
\def\Mdyn{\mbox{$M_{\rm dyn}$}}

\def\Mket{\mbox{$M_{\rm dust}^{K=\,3}$}}

\def\td{\mbox{$t_{\rm d}$}}
\def\Fco{\mbox{$F_{\rm co}$}}

\def\Llobe{\mbox{$l_{\rm lobe}$}}
\def\Vt{\mbox{$V_{\rm t}$}}
\def\Vb{\mbox{$V_{\rm b}$}}
\def\Vflow{\mbox{$V_{\rm flow}$}}
\def\Mlobe{\mbox{$M_{\rm lobe}$}}

\def\Tk{\mbox{$T_{\rm k}$}}
\def\Trot{\mbox{$T_{\rm rot}$}}

\def\Tsb{\mbox{$T_{\rm sb}$}}

\def\Tex{\mbox{$T_{\rm ex}$}}
\def\Tbg{\mbox{$T_{\rm bg}$}}
\def\Tdust{\mbox{$T_{\rm dust}$}}

\def\NNtwoH{\mbox{$N_{\rm N_2H^+}$}}

\def\NMCN{\mbox{$N_{\rm CH_3CN}$}}
\def\meanNmol{\mbox{$\langle N$(H$_2)\rangle$}}

\def\taut{\mbox{$\tau_{\rm tot}$}}

\def\gs{G\,16.59$-$0.05}
\def\gtt{G\,23.01$-$0.41}
\def\gte{G\,28.87$+$0.07}
\def\lesssim{\mathrel{\hbox{\rlap{\hbox{\lower4pt\hbox{$\sim$}}}\hbox{$<$}}}}
\def\gtrsim{\mathrel{\hbox{\rlap{\hbox{\lower4pt\hbox{$\sim$}}}\hbox{$>$}}}}
\def\Req{\mbox{$R_{\rm eq}$}}
\def\veq{\mbox{$V_{\rm eq}$}}
\def\Reff{\mbox{$R_{\rm eff}$}}
\def\vrot{\mbox{$V_{\rm rot}$}}
\def\Mdyn{\mbox{$M_{\rm dyn}$}}
\def\Mdk{\mbox{$M_{\rm dust}^{K=3}$}}

\slugcomment{Accepted to ApJ on 2007 September 17}

\shorttitle{Rotating Toroids around High-Mass (Proto)Stars}
\shortauthors{R. S. Furuya et al.}

\begin{document}


\title{Candidate Rotating Toroids around High-Mass (Proto)Stars}


\author{R. S. Furuya\altaffilmark{1}}
\affil{Subaru Telescope, National Astronomical Observatory of Japan}
\email{rsf@subaru.naoj.org}

\author{R. Cesaroni\altaffilmark{2}}
\affil{INAF, Osservatorio Astrofisico di Arcetri}
\email{cesa@arcetri.astro.it}

\author{S. Takahashi\altaffilmark{3}}
\affil{Department of Astronomical Science, Graduate University for Advanced Studies}
\email{satoko\_t@asiaa.sinica.edu.tw}

\author{C. Codella\altaffilmark{2}}
\affil{INAF -- Istituto di Radioastronomia, Sezione di Firenze}
\email{codella@arcetri.astro.it}

\author{M. Momose\altaffilmark{4}}
\affil{Institute of Astronomy and Planetary Science, Ibaraki University}
\email{momose@mx.ibaraki.ac.jp}

\and

\author{M. T. Beltr\'an\altaffilmark{5}}
\affil{Department d'Astronomia i Meteorologia, Universitat de Barcelona}
\email{mbeltran@am.ub.es}

\altaffiltext{1}{650 North A'ohoku Place, Hilo, HI 96720}
\altaffiltext{2}{Largo Enrico Fermi 5, I-50125 Firenze, Italy}
\altaffiltext{3}{Current address: Academia Sinica Institute of Astronomy and Astrophysics, P.O. Box 23-141, Taipei 106, Taiwan, R.O.C.}
\altaffiltext{5}{Bunkyo 2-1-1, Mito, Ibaraki 310-8512, Japan}
\altaffiltext{6}{Av. Diagonal, 647, 08028, Barcelona, Catalunya, Spain}


\begin{abstract}
Using the OVRO, Nobeyama, and IRAM mm-arrays, we searched for 
``disk''-outflow systems in three high-mass (proto)star forming regions:  
\gs, \gtt, and \gte.
These were selected from a sample of \amm\ cores (Codella, Testi \& Cesaroni) 
associated with OH and \wat\ maser emission (Foster \& Caswell) and with no
or very faint continuum emission.  
Our imaging of molecular line (including rotational transitions of \mcn)
and 3\,mm dust continuum emission revealed that these are 
compact ($\lesssim$ 0.05 -- 0.3 pc),
massive ($\sim$ 100 -- 400 \Msun), and
hot ($\sim$100 K) molecular cores (HMCs), that is
likely sites of high-mass star formation prior to the appearance
of ultracompact \hii\ regions.
All three sources turn out to be associated with molecular outflows
from \co\ and/or HCO$^+$ $J=$1--0 line imaging.
In addition, velocity gradients of 10 -- 100 \kms\ pc$^{-1}$
in the innermost ($\lesssim$0.03 -- 0.13 pc),
densest regions of the \gtt\ and \gte\ HMCs are identified
along directions roughly perpendicular to the axes 
of the corresponding outflows. 
All the results suggest that these cores might be rotating about the outflow axis, 
although the contribution of rotation to gravitational equilibrium of the HMCs
appears to be negligible. 
Our analysis indicates that the 3 HMCs are close to virial equilibrium due to
turbulent pressure support.
Comparison with other similar objects where rotating toroids have 
been identified so far shows that in our case rotation appears to be much less
prominent; this can be explained by the combined effect of 
unfavorable projection, large distance, and limited angular resolution
with the current interferometers.
\end{abstract}





\section{Introduction}
\label{ss:intro}

The role of disks in the formation process of low-mass stars (i.e. stars with
masses $\lesssim 1$ \Msun) has been extensively studied in the last two decades
through high angular resolution observations at various wavelengths.  Images
of such disks have been obtained in the optical (e.g. Burrows et al. 1996)
and at mm-wavelengths (e.g. Simon et al. 2000).  The latter have demonstrated
that the majority of the disks undergo Keplerian rotation.  These findings are
consistent with the fact that low-mass stars form through accretion while
(partial) conservation of angular momentum during the dynamical collapse
produces a flattened and rotating structure at the center of the core.\par

What about high-mass ($M_\ast \gtrsim 8$ \Msun) stars?  In this case,
formation through accretion faces the problem that stars more massive than
$\sim$8~$M_\odot$ reach the zero age main sequence still deeply embedded in
their parental cores (Palla \& Stahler 1993).  At this point radiation
pressure from the newly formed early-type star can halt the infall, thus
preventing further growth of the stellar mass.  Various solutions have been
proposed to solve this problem:
(i)~massive stars might form through merging of lower mass stars (Bonnell,
Bate \& Zinnecker, H. 1998; Bonnell \& Bate 2002; Bally \& Zinnecker 2005);
(ii)~sufficiently large accretion rates could allow the ram pressure of the
infalling material to overcome the radiation pressure form the star
(Behrend \& Maeder 2001; McKee \& Tan 2003; Bonnell, Vine, \& Bate 2004)
(iii)~non-spherical accretion could weaken the effect of radiation pressure
by allowing part of the photons to escape through evacuated regions along
the outflow axis and, at the same time, enhancing the ram pressure of the
accreting material by focusing it through the disk plane (Yorke \&
Sonnhalter 2002; Krumholz, McKee, \& Klein 2005).\par

An important test to discriminate between the different hypotheses is the
presence of rotating, circumstellar disks, which would lend support to the
third scenario depicted above. This can be determined by
inspecting the velocity field of the innermost parts of 
molecular cores where young massive (proto)stars are believed to form.
One possibility is to look for velocity gradients perpendicular to the
direction of the larger scale outflows associated with such cores; 
this would suggest that one is observing rotation about the outflow axis.
This technique has been adopted by us and other authors successfully
inferring the existence of rotation in the gas enshrouding high-mass young
stellar objects (YSOs) and leading to the discovery of circumstellar
disk-like structure or ``toroids'' in a limited number of cases 
(see Cesaroni et al. 2007 for a review on this topic).\par

The goal of the present study was to establish whether the presence of
rotation is common in high-mass star forming regions by observing three more
objects, expected to be sites of deeply embedded OB (proto)stars.  The ideal
target for this type of studies are hot molecular cores (HMCs),
which are believed to be the cradles
of massive stars (see e.g. Kurtz et al. 2000; Cesaroni et al. 2007).  Beside
other studies, the one by Codella, Testi \& Cesaroni (1997, hereafter CTC97),
has proved successful in identifying the birthplaces of young massive stars
as dense \amm\ cores associated with OH and H$_2$O maser emission.  One of
these HMCs (G\,24.78+0.08) has been the subject of a series of articles by us
(Furuya et al. 2002; Cesaroni et al. 2003; Beltr\'an et al. 2004, 2005),
which have shown that this is a unique object, characterized by the
simultaneous presence of rotation, outflow, and infall towards a hypercompact
\hii\ region ionized by an O9.5 star (Beltr\'an et al. 2006). 
With this in mind, we completed the study of the sources in CTC97 by observing
three more objects in the same tracers used to investigate G\,24.78+0.08.  The
selected targets for this study are G\,16.59$-$0.05 at $d =$ 4.7 kpc, 
G\,28.87$+$0.07 at 7.4 kpc, and G\,23.01$-$0.41 at 10.7 kpc.
The fact that in all three cases no or only faint free-free continuum
emission has been detected, notwithstanding the large luminosities of the
sources (CTC97), suggests that the embedded YSOs might be in an
even earlier evolutionary phase than those in G\,24.78+0.08 
(Furuya et al. 2002; Beltr\'an et al. 2004).

\section{Observations and Data Retrieval}
\label{ss:obs}

\subsection{Millimeter Array Observations}
\label{ss:mmarray}

Aperture synthesis observations of molecular lines and continuum emission at
3\,mm were carried out using the Owens Valley Radio Observatory
(OVRO)\footnote{Research at the Owens Valley Radio Observatory is supported
by the National Science Foundation through NSF grant AST 02-28955.}
millimeter array towards \gs\ and \gte, and the Nobeyama Millimeter Array
(NMA) of the Nobeyama Radio Observatory\footnote{Nobeyama Radio Observatory
is a branch of the National Astronomical Observatory, operated by the
Ministry of Education, Culture, Sports, Science and Technology, Japan.}
towards \gtt.  We also carried out \co\ (1--0) line and millimeter continuum
emission imaging with the Plateau de Bure Interferometer (PdBI) of the
Institut de Radio Astronomie Millim\'etrique\footnote{IRAM is supported by
INSU/CNRS (France), MPG (Germany), and IGN (Spain).} (IRAM).  The parameters
for the continuum and molecular line observations are summarized in Tables
\ref{tbl:obs_cont} and \ref{tbl:obs_lin}, respectively.

\subsubsection{OVRO Observations}
\label{ss:ovro}

The OVRO observations of \gs\ and \gte\  were carried out in the period from
2003 September to 2004 May in 3 array configurations (E, H, and UH).  The
shortest projected baseline length, i.e. the shadowing limit, is about
12.8\,m.  This makes our OVRO observations insensitive to structures more
extended than $54\arcsec$, corresponding to 1.2 and 1.9~pc at the distances
of \gs\ and \gte, respectively.  We observed the \mcn\ (5--4) and
\NtwoH\ (1--0) lines in the upper sideband (USB) and the \HCO\ (1--0) line in
the lower sideband (LSB).  For the continuum emission, we simultaneously used
the Continuum Correlator with an effective bandwidth of 4 GHz and the newly
installed COBRA with 8 GHz bandwidth.  We configured the digital correlator
with 31\,MHz bandwidth and 62 channels centered at 91979.970 MHz, so to cover
the $K=$ 0 to 3 components of the \mcn\ (5--4) transition.  We used 3C\,273
and 3C\,454.3 as passband calibrators, and NRAO\,530 and J1743$-$038 as phase
and gain calibrators respectively for \gs\ and \gte.  The flux densities of
NRAO\,530 and J1743$-$038 were determined from observations of Uranus and
Neptune.  We estimate the uncertainty of flux calibrations to be 10\%.  The
data were calibrated and edited using the MMA and MIRIAD packages.  We
constructed continuum images from the COBRA data by employing the
multi-frequency synthesis method:  the final center frequency is 90.584\,GHz.

\subsubsection{NMA Observations}
\label{ss:nma}

Our NMA observations towards \gtt\ were carried out in the period from 2003
December to 2004 May with 3 array configurations (D, C, and AB).  The
largest detectable source size is $50\arcsec$, corresponding to 2.6 pc at
$d~=$ 10.7 kpc.  We observed the \mcn\ (6--5), HNCO (5--4), \tCO\ (1--0) and
\CeO\ (1--0) lines in the USB.  For the continuum and line emission, except
for \mcn, we employed the Ultra Wide Band Correlator (UWBC)
with a 512\,MHz bandwidth in each sideband;
this configuration gives a total bandwidth of 1\,GHz for the continuum and a
velocity resolution of 5.5 \kms\ for the lines.  We used the FX correlator
with 32\,MHz bandwidth centered at 110374.0\,MHz, thus covering the $K=$
0 to 3 components of the \mcn\ (6--5) transition.  We adopted 3C\,273 as
passband calibrator and J1743$-$038 as phase and gain calibrator.  The flux
densities of J1743$-$038 were bootstrapped from Uranus, and the uncertainty
of the flux calibration is estimated to be 10\%.  We decided not to use the
\tCO\ and \CeO\ data taken with the C and AB configurations as these are too
extended and resolve the emission.  All the data were calibrated and edited
using the UVPROC2 and MIRIAD packages.

\subsubsection{PdBI Observations}
\label{ss:pdbi}

Our \co\ (1--0) and 2.6\,mm continuum emission observations towards the 3 HMCs
were carried out with the IRAM 5-antenna interferometer on Plateau de Bure
in 1998 April and May. 
The D and C1 configurations were used, yielding a largest detectable
angular scale of $\sim$\,44\arcsec.
The 82 -- 116 GHz SIS receivers were tuned in SSB at the frequency of the
\co\ (1--0) line, and
the facility correlator was configured with a bandwidth of 40\,MHz 
centered at the same frequency.
Continuum emission has also been 
measured with two bandwidths of 160\,MHz each. Line free channels were
averaged to produce a continuum image which was then subtracted from the line
data in the visibility plane.
The flux densities of J1833$-$210 (1.7\,Jy) and J1743$-$038 (3.0\,Jy)
were bootstrapped from 3C\,273 whose flux density was assumed to be 16.7\,Jy.
We estimated an overall uncertainty on flux calibration of 20\%. 
Data editing, calibrations, and image construction 
were done using the GILDAS software package developed at IRAM. 
Tables \ref{tbl:obs_cont} and \ref{tbl:obs_lin} summarize the observing
parameters for the PdBI continuum and line imaging, respectively.

\subsection{GLIMPSE and MSX Archive Data}
\label{ss:ircont}

We have retrieved infrared (IR) images at 21 \mic\ from the Midcourse Space
Experiment (MSX; Price et al. 2001) survey and at 8.0, 5.8, 4.5, and 3.6 \mic\
from the Galactic Legacy Infrared Mid-Plane Survey (GLIMPSE) survey making
use of the Infrared Array Camera (IRAC) on-board the Spitzer satellite.  The
calibrated data from the {\it Spitzer Science Center} were processed through
the GLIMPSE pipeline reduction system (Benjamin et al. 2003; Whitney et al.
2004).  All the infrared (IR) images were used without spatial smoothing; the
pixel size is 6\arcsec\ and $\sim 1\farcs 2$ for the MSX and IRAC data,
respectively.

\section{Results and Discussion}
\subsection{Continuum Emission}
\label{s:res_cont}
\subsubsection{Source Identification Based on Millimeter Images}
\label{ss:contmaps}

\begin{figure}[t]
\begin{center}
\includegraphics[width=8.6cm,angle=0]{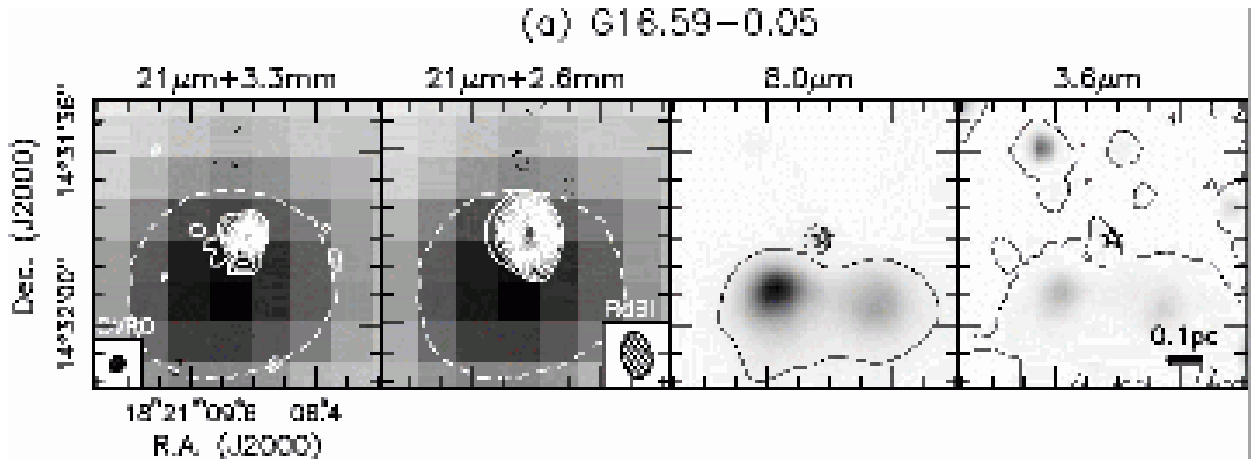} \\
\includegraphics[width=8.6cm,angle=0]{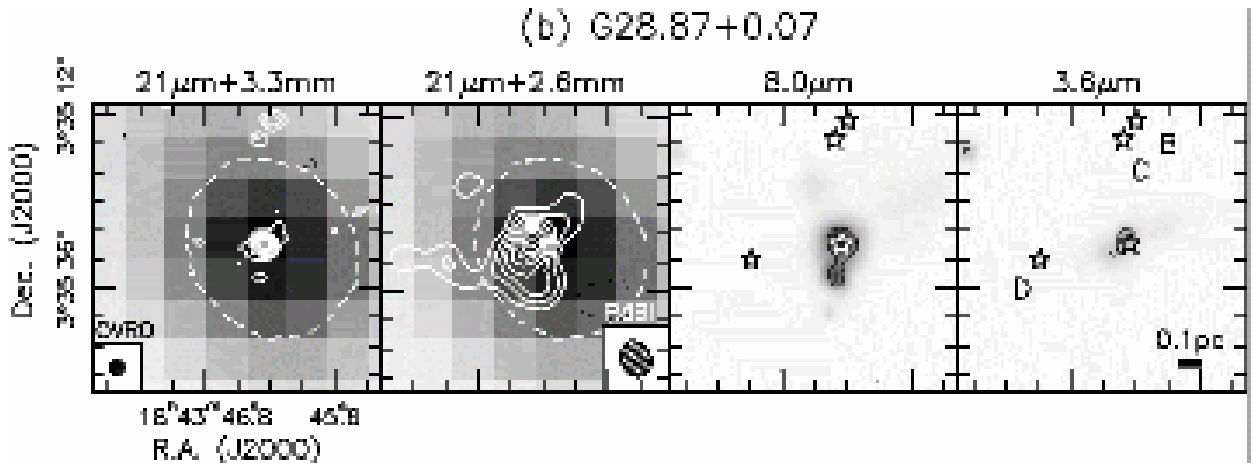} \\
\includegraphics[width=8.6cm,angle=0]{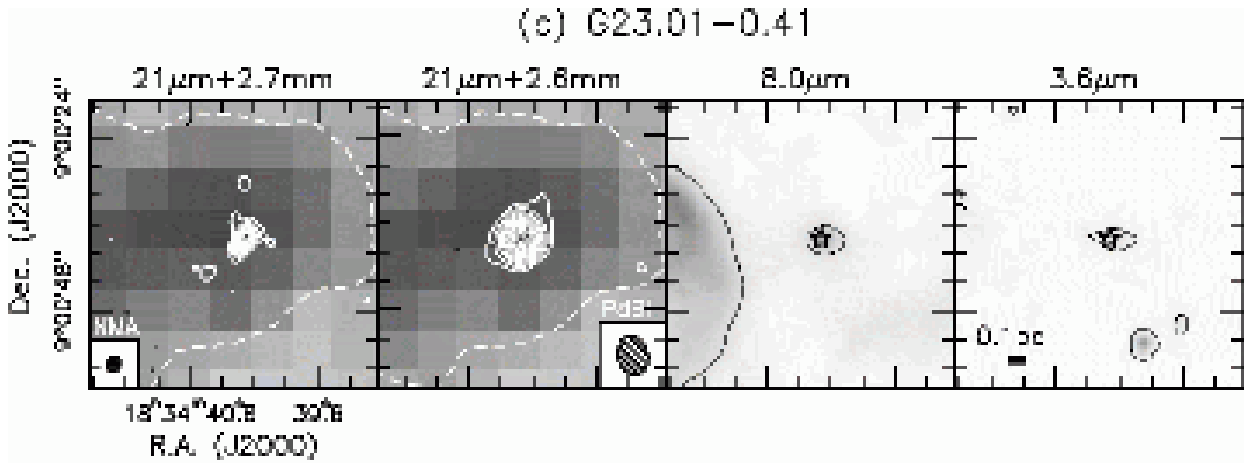}
\end{center}
\caption{\small Continuum emission maps towards the 3 HMCs at mm- and infrared wavelengths.
The left two panels for each source shows an overlay of mm-continuum emission
(white contours) on the 21 \mic\ one from the MSX satellite data base;
the single white dashed contour indicates the 50\% level of the 21 \mic\ emission.
The ellipses in the inserted panels show the HPBWs of the synthesized beam 
(Table \ref{tbl:obs_cont}).
The right two panels show 8.0 and 3.6 \mic\ images from
the Spitzer GLIMPSE survey; the stars mark the peak positions of
the mm continuum source, and the thin contour indicate the 50\% level contour
of the peak intensity of the main cores of our interests.
Contours for the mm-continuum maps are 2$\sigma$ intervals, 
starting from the 3$\sigma$ level
(see Table \ref{tbl:obs_cont} for the RMS noise levels).
In the panel (b), the associated labels indicate names of the mm sources identified
by us (see $\S\ref{ss:contmaps}$ and Table \ref{tbl:res_cont}).
}
\label{fig:contmaps}
\end{figure}

Figure \ref{fig:contmaps} compares our 3\,mm continuum emission maps to the
IR images. The latter correspond to the MSX data at 21 \mic\ and the GLIMPSE
ones at the longest (8.0 \mic) and shortest (3.6 \mic) wavelengths.  Here we
do not show GLIMPSE images at 5.8 and 4.5 \mic\ as they are basically similar
to those at 8.0 and 3.6 \mic.  All 3 objects show intense, compact 3\,mm
continuum emission coincident with IR emission, although one must consider
that the angular resolution of the MSX images is much worse than that of our
interferometric maps. 
In the millimeter maps of \gte, one can also see weaker 3.3\,mm continuum
sources located $\simeq$ 14\arcsec\ to the north 
(hereafter \gte B and C; see Figure \ref{fig:contmaps}b) 
and a 2.6\,mm source $\simeq$ 10\arcsec\ to the east (\gte D; same figure) 
of the main core found by CTC97. 
No other mm continuum emission is detected with
S/N~$\ge$~5 in our OVRO, NMA, and PdBI fields of view towards the 3 sources.
Assuming that the flux density \Snu, varies as $\nu^{(2+\beta)}$
with $\beta=1.5$ (Preibisch 1993; Molinari et al. 2000), the expected flux
of \gte~B and~C in the PdBI beam at 2.6~mm is below the sensitivity of
our images (3.5~mJy corresponding to 2.2$\sigma$). Similarly, assuming that
\gte~D matches the PdBI beam, this source is resolved by the OVRO beam and
its flux at 3.3~mm corresponds to only 1.4$\sigma$ of the OVRO continuum image,
thus explaining why it is not detected.

The GLIMPSE images of \gs\ and \gte\ show IR emission both 
from the main cores and from their surroundings.
On the other hand, \gtt\ seems rather isolated, although we may be missing
weaker objects due to the large distance to the region.
In the following we make no attempt to identify other IR sources
beside the counterparts of those detected at mm wavelengths, because
this would require an analysis which goes beyond the purposes of the
present study.
Notice that no other mm sources with 3$\leq$ S/N $<$ 5 is seen
in the IRAC bands.
We summarized the identified continuum sources in Table \ref{tbl:res_cont},
where peak positions and effective radius (\Reff ) of the mm-continuum sources as well
as their $S_\nu$ integrated inside the $5\sigma$ level contours are given. 
Here the observed angular radius of the source is computed from
$\sqrt{\cal{A}/\pi}$ where $\cal{A}$ is the area enclosed by the 50\%
contour level of the emission. Deconvolution assuming the source and the
synthesized beam to be Gaussian is then applied, thus obtaining \Reff, 
assuming that the emission has a Gaussian distribution.
It should be noted that \Reff\ does not depend on the image noise levels. 
To estimate the core mass from the continuum flux the latter has been
integrated all over the emitting region,
namely inside the corresponding 5$\sigma$ contour level.
The fluxes of the IR counterparts are listed in Table \ref{tbl:IRfluxes}.\par

\subsubsection{Continuum Spectra and Core Masses Estimated from the 3\,mm Flux Densities}
\label{ss:sed}

\begin{figure}
\begin{center}
\includegraphics[width=.35\linewidth,angle=-90]{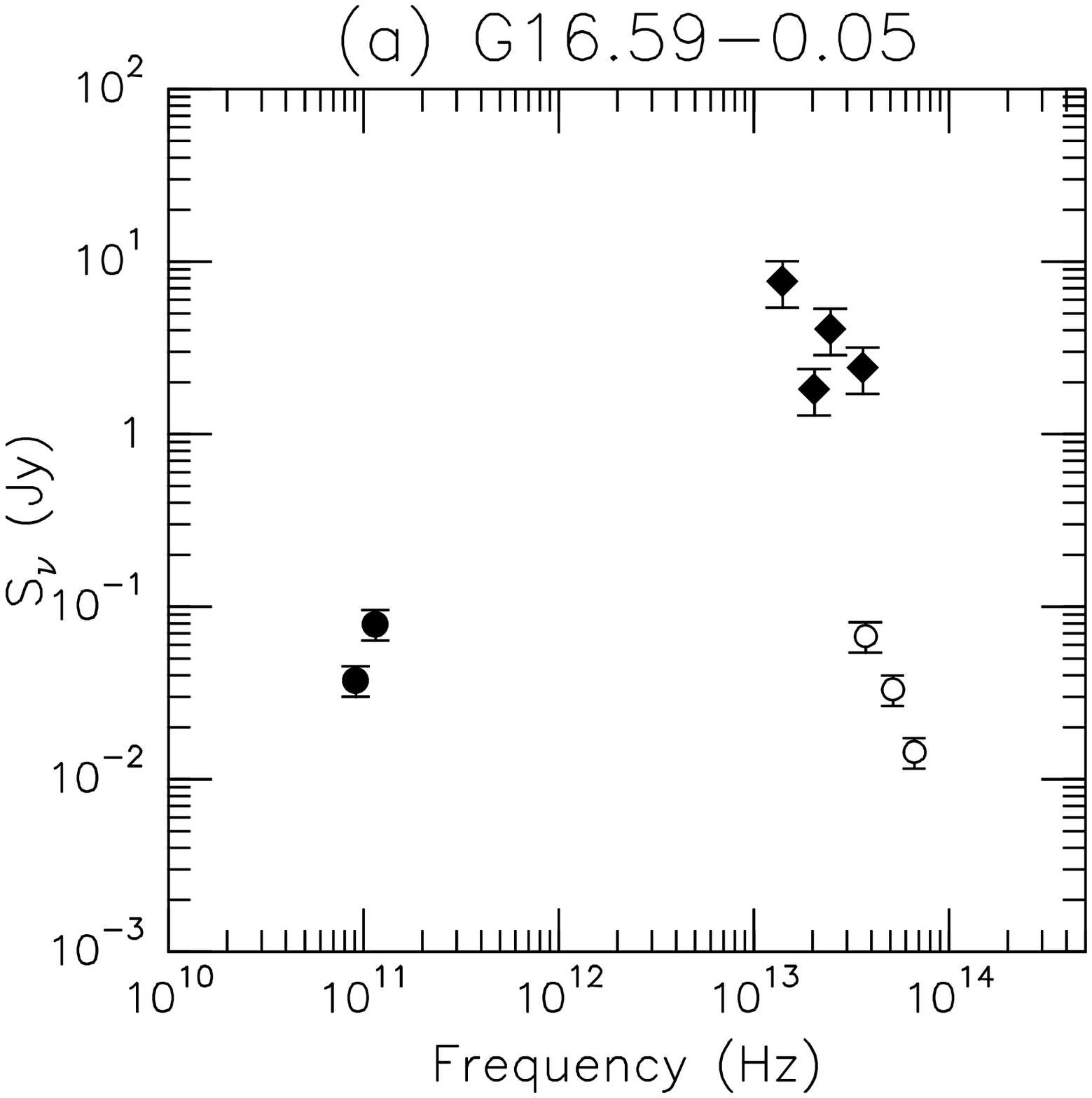}
\includegraphics[width=.35\linewidth,angle=-90]{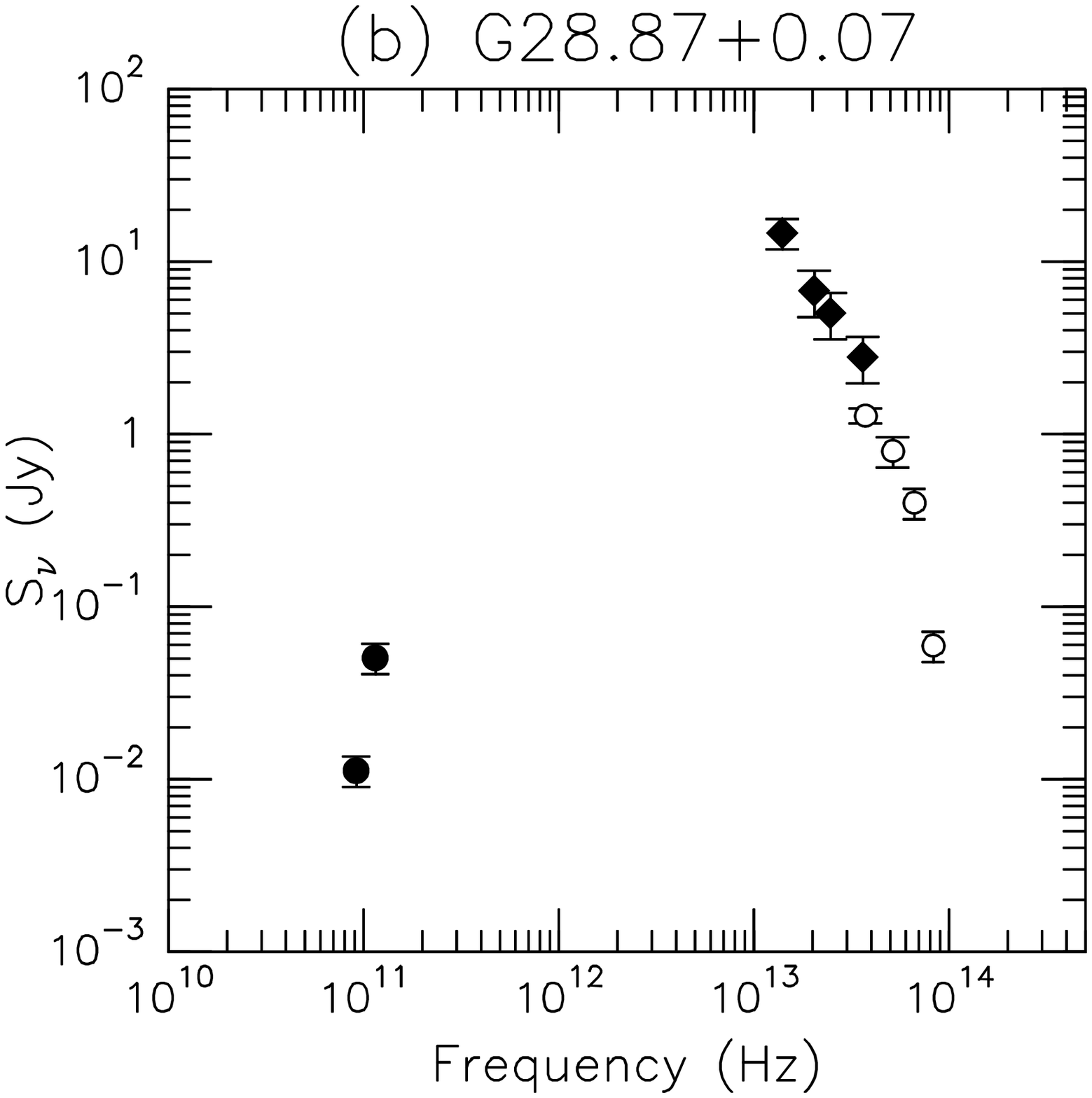}
\includegraphics[width=.35\linewidth,angle=-90]{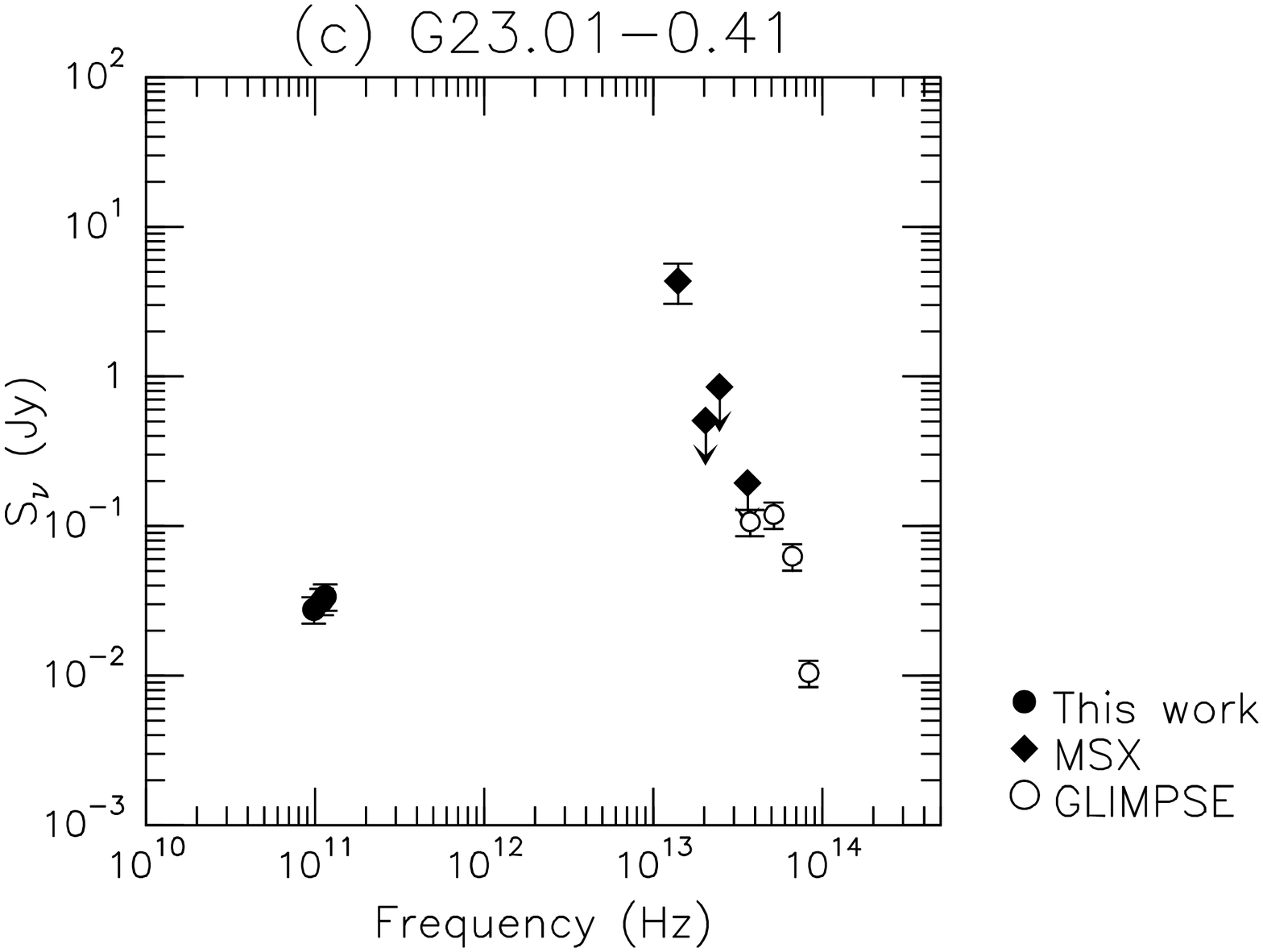}
\end{center}
\caption{Continuum spectra of the 3 HMCs. 
The flux densities at 3\,mm and infrared bands are summarized 
in Tables \ref{tbl:res_cont} and \ref{tbl:IRfluxes}, respectively.
}
\label{fig:sed}
\end{figure}

Figure \ref{fig:sed} shows the continuum spectra of the 3 HMCs, as from
Tables \ref{tbl:res_cont} and \ref{tbl:IRfluxes}.
No contamination by free-free emission is expected to affect
our 3~mm fluxes, as the upper limits obtained by CTC97 at 1.26\,cm guarantee
a maximum free-free flux of $\simeq$ 1.4\,mJy at 3.0\,mm -- assuming optically
thin emission.
Establishing the origin of the IR emission requires a model fitting of the
spectral energy distribution (SED) that goes beyond the purpose of this study.
We only note that a simple grey-body fit cannot consistently reproduce the SED
from the mm to the near-IR regime. This suggests that 
the structure of the cores is rather complex, probably hosting a multiple stellar system 
such as the one seen in the G\,29.96$-$0.02 star forming region: here,
recent sub-arcsecond resolution 
($\sim$ 1800\,AU resolution in linear scale) Submillimeter Array (SMA) imaging resolved 
the HMC (Maxia et al. 2001; Olmi et al. 2003) into six submm continuum sources
(Beuther et al. 2007).
Although our angular resolutions in AU are 5--12 times worse, 
by analogy to the case of G\,29.96$-$0.02, we may argue that 
multiple sources are present in our cores, which might explain why the SEDs cannot
be fitted with a simple grey-body fit.
Furthermore, temperature gradients as well as optical depth
effects and clumpiness should be taken into account to obtain a satisfactory fit.
In particular images with arcsecond resolution at wavelengths above 20~$\mu$m
would be of great importance to obtain a precise estimate of the bolometric
luminosity (and hence of the mass) of the embedded (proto)stars.\par

The 3\,mm flux density in Table \ref{tbl:res_cont} can be used to estimate
the mass of the cores (\Mdust), under a few assumptions on the dust
emissivity.  More specifically, we use Hildebrand's (1983) relation \Mdust
$=\frac{S_\nu d^2}{\kappa_\nu B_\nu (T_{\rm dust})}$ with $\kappa_\nu = 5.0\times
10^{-3} (\nu/231\,{\rm GHz})^\beta$ cm$^2$ g$^{-1}$ and $\beta =1.5$.
When calculating \Mdust, we set \Tdust\ equal to the rotational temperature 
obtained from the \mcn\ lines (see $\S$\ref{ss:rd}), 
assuming that gas and dust are well-coupled
(e.g., Kr\"ugel \& Walmsley 1984), a reasonable
hypothesis at densities as high as those traced by the \mcn\ emission
(typically $>10^{6-7}$ \cmc).\par

The derived values of \Mdust\ range from 95 to 380 \Msun\ (Table~\ref{tbl:Mcore}).  
Our 3 targets appear to fall in the mass range for ``heavy'' 
HMCs ($\gtrsim 100$ \Msun; Cesaroni et al. 2007).
Cores that massive are likely to host multiple stars,
as opposed to ``light'' HMCs ($<$ a few 10 \Msun),
which appear to contain only single OB stars or binary systems 
(Cesaroni et al. 2007).\par

In \gte\ we have identified 3 additional weaker sources.
Assuming that \Tdust\ is equal to the excitation temperature of \NtwoH(1--0)
(see $\S$\ref{ss:totmaps-gte} and Figure \ref{fig:gte-maps}c),
we obtain \Mdust $=$ 7\Msun\ with \Tdust $=$ 46\,K for \gte D.
The excitation temperature is estimated from the hyperfine structure 
analysis of mean spectra of the \NtwoH\ emission towards a
condensation hosting \gte D.
If we assume that \Tdust $\simeq$ 46\,K is valid for \gte B and C,
we obtain \Mdust\ of 29 \Msun\ for \gte B and 28 \Msun\ for \gte C.
These results suggest that the \gte\ HMC
(\Mdust $=$ 100\Msun; Table \ref{tbl:Mcore}) is surrounded by less massive objects.  
Such a situation appears to be
typical of high-mass star forming regions, where embedded OB (proto)stars
happen to be surrounded by less embedded, possibly more evolved, lower mass stars
(Molinari et al. 1998; Fontani et al. 2004a, 2004b).
Furthermore, because of the absence of free-free emission (CTC97),
we argue that the \gte\ cluster is much younger than
other clusters of massive YSOs, such as, e.g., the cluster of ultracompact
\hii\ regions in G\,19.61$-$0.23 (Furuya et al. 2005).\par


Finally, in order to estimate the \mcn\ abundance in the cores (see
$\S$\ref{ss:Xmcn}), we have calculated the mean molecular hydrogen column
density, \meanNmol, over the cores.  To make the comparison between
continuum and \mcn\ emission as consistent as possible, we reconstructed the
3\,mm images by fixing the beam sizes to those of the \mcn\ images 
(Table \ref{tbl:obs_lin}). 
We subsequently integrated the continuum flux over the
region enclosed by the 5\sgm\ level contours.
In Table~\ref{tbl:Mcore} we summarize the calculated mean column densities 
(\meanNmol) which are (3 -- 4)$\times10^{23}$ \cmq, 
implying optical depths on the order of $10^{-3}$,
consistent with our assumption that the dust emission is optically thin at 3\,mm.

\subsection{Spectral Line Profiles towards the Cores}
\label{ss:sp}

Figure \ref{fig:sp} shows the line spectra at the peak positions of each
integrated intensity map, in units of brightness temperature in the
synthesized beam (\Tsb ).  For the \mcn\ emission, we show the $K=3$
component as this is not blended with the other $K$-ladder lines.\par

\paragraph{G\,16.59$-$0.06}
The \co\ and \HCO\ peak spectra present deep ``absorption'' features around the
systemic velocity, the latter being estimated from the \mcn\ lines (see
$\S\ref{ss:rd}$).  Most likely this is not due to real self-absorption but to
the fact that emission at low velocities arises from extended regions
resolved out by the interferometer, as found in the case of G24.78+0.08 (see
Cesaroni et al. 2003).  However, only single-dish mapping of the regions
under study in the same lines can establish the nature of such absorption.\par

The \co\ spectrum shows prominent high-velocity wings on both sides of the
line, indicating the presence of a molecular outflow.
In order to adopt suitable velocity ranges for the outflow lobes
traced by the \co\ emission, we define terminal (\Vt) and boundary (\Vb)
velocities:  \Vt\ is the most blue- and redshifted LSR-velocity where the wing
emission drops below the 1.5$\sigma$ level and \Vb's are the velocities where
the intensity of the \mcn\ $K = 3$ line falls below 1.5$\sigma$ 
(see caption of Figure \ref{fig:gs-maps} for the \Vt\ and \Vb\ values).

\paragraph{G\,28.87$+$0.08}
The \co\ and \HCO\ (1--0) spectra towards \gte\ are very similar each other
in the sense that the emission around \Vsys\ is almost missing, and the
blueshifted wing is barely detected, whereas prominent redshifted emission
is seen.  The \mcn\ $K=3$ emission peaks around \Vlsr\ $\simeq 103$ \kms\
where the \co\ and \HCO\ spectra present a sharp edge.  Also in this case, as
for \gs, we argue that the spectral profiles of the \co\ and \HCO\ lines may
suffer of missing short spacing in our interferometric observations.

\paragraph{G\,23.01$-$0.41}
The profile of the \co\ line towards \gtt\ presents a
redshifted wing much more prominent than
the blue one, while the emission close to \Vsys\ is strongly depressed.
On the other hand, the lines of the lower density tracers \tCO\ and \CeO\ present
well defined peaks, although the velocity resolutions are poor.
The peak LSR-velocities of the \tCO\ and \CeO\ emission shifted slightly 
towards the blue with respect to
the peak velocity of \mcn\ -- which we have assumed as representative of the
bulk material.

\subsection{Maps of the Molecular Line Emission}
\label{ss:totmaps}
\subsubsection{G\,16.59$-$0.06}
\label{ss:totmaps-gs}

Figure \ref{fig:gs-maps} compares maps of molecular line and continuum
emission in \gs.  Panel (a) shows an overlay of the \co\ wing emission on the
3\,mm continuum emission.  The wing emission is integrated over the velocity
ranges between \Vb\ and \Vt\ for both the blue and red lobes ($\S$\ref{ss:sp}).  
The \gs\ outflow map shows a clear bipolarity along a line with P.A.$\simeq$135\degr.
We define the outflow axis as the line connecting the peaks of the blue lobe
and 3\,mm continuum map.  The blue lobe shows a remarkable structure
with a large opening angle of $\simeq 180^{\circ}$, whereas the red lobe is
much more collimated.  Interestingly, both lobes have a large aspect ratio
($>1$), defined as the ratio between the width and the length of the lobe.
Moreover, the two lobes have a rather large spatial overlap, coinciding also
with the position of the 3\,mm continuum emission.  
Such a geometry indicates that the inclination angle of the outflow axis 
with respect to the line-of-sight -- albeit difficult to quantify precisely
-- must be rather small.\par

In Figure \ref{fig:gs-maps}d, we see that, while the distributions of \mcn\
(5--4) $K = 0+1$ and \amm\ (3,3) emission agree well with each other, 
the peak position of the 3.3\,mm
source is displaced by $\simeq 1\farcs4$ to the northwest of the \mcn\ peak.
We also verified that the peak positions of the $K = 2$ and 3 emission agree with
that of the $K=0+1$ emission.
Note that the 3.3\,mm continuum and \mcn\ visibilities shared the same
calibrators as they were observed simultaneously ($\S\ref{ss:ovro}$), 
and that the peak positions of the 3.3 and 2.6 \,mm continuum sources
(see Table \ref{tbl:res_cont}), measured respectively with OVRO and PdBI, 
agree with each other within the errors. 
These facts indicate that the positional displacement between 
the \mcn\ and 3\,mm continuum emission is real 
(see $\S\ref{ss:VelSt}$ for a discussion).
One cannot rule out the possibility that such a displacement is due
to optical depth effects, although this seems unlikely as line emission
peaks at the same position for all species (\amm\ and \mcn), and the \mcn\ opacity
quite low ($\lesssim2$; described in $\S\ref{ss:rd}$).
Observation of higher-excitation \mcn\ lines might help shedding light on this issue.
The peak position of the more extended \HCO\ emission agrees with those of
\amm\ and \mcn, suggesting that the \HCO\ (1--0) line also traces the HMC.  On
the other hand, the \NtwoH\ emission has a totally different morphology from
the other molecules, in the sense that it has a fairly extended structure and
its peak position differs from both those of the other lines and that of the
3\,mm continuum.  The \NtwoH\ emission is more extended than the \HCO\
emission, suggesting that it might be associated not only with the HMC, but
also with the other sources seen in the IR images (Figure \ref{fig:contmaps}a).\par

To estimate the column density of the region, we analyzed the
\HCO\ and \NtwoH\ data in the following fashion.  The mean column density of
a linear rigid rotor is calculated from Eq.~(A1) of Scoville et al. (1986).
Since we do not have measurements of the optical depth ($\tau$) and
excitation temperature (\Tex) of the \HCO\ line, we assumed that 
the \HCO\ emission is
optically thin and estimated the column density for two extreme values \Tex:
a minimum \Tex, obtained from the line peak \Tsb\ (32.8\,K from Figure
\ref{fig:sp}a) and a maximum \Tex, from \Trot\ of the \mcn\ emission (130\,K;
described in $\S\ref{ss:rd}$).  Correspondingly, the mean \HCO\ column
density over a region enclosed by its 5$\sigma$ contour level is $N_{\rm
HCO^+}\simeq6.3\times 10^{14}$ \cmq\ and $2.1\times 10^{15}$ \cmq.  For the
fractional abundance, $X$(\HCO), we assume a value of $10^{-9}$ (see e.g. van
Dishoeck et al. 1993).  Under this assumption, the mass traced by the
\HCO\ (1--0) line (\Mlte) ranges from 440 to 1500 \Msun\ (Table
\ref{tbl:clumps}).  We stress that these estimates are to be taken as lower
limits, as part of the line emission close to the systemic velocity is
resolved out in our interferometric maps.\par

Subsequently we analyzed the \NtwoH\ data because the hyperfine structure
(HFS) of the transition allows an estimate of the line optical depth (see,
e.g., Benson, Caselli \& Myers 1998; Caselli et al. 2002).  
The details of our \NtwoH\ HFS analysis are described in Appendix~B of
Furuya et al. (2006).
For this purpose,
we used two spectra obtained integrating the emission over the 5$\sigma$
contour level of the 3\,mm continuum emission and the integrated \NtwoH\ line
emission.  Note that the \NtwoH\ emission does not peak at the position of
the 3\,mm continuum source. For the first spectrum, our analysis gives $\Tex
\simeq 41\pm9$~K, total optical depth (corresponding to the sum of the
optical depths of the 7 hyperfine components) $\taut\simeq 0.28\pm 0.03$,
and intrinsic line width $\dVint\simeq 2.10\pm0.04$ \kms\ towards the 3\,mm
source.  From the second spectrum, we obtain \Tex\ $=$ 16$\pm$2 K, \taut\ $=$
0.91 and \dVint\ $=$ 2.82$\pm$0.02 \kms.  Using Eq.~(B7) of Furuya et al.
(2006), we calculate a column density $\NNtwoH \simeq (7.8\pm 1.5)\times
10^{13}$ \cmq\ towards the 3\,mm continuum source and $(6.1\pm 0.4)\times
10^{13}$ \cmq\ for the average spectrum.  If we assume X(\NtwoH) $= 3\times
10^{-10}$ suitable for high-mass star forming regions (Womack et al. 1992),
the mass estimate for the entire region is 540 \Msun\ (Table
\ref{tbl:clumps}).\par

These numbers demonstrate that the \NtwoH\ line is coming from a lower (column)
density region, distinct from the HMC traced by the continuum and \mcn\ line
emission -- whose analysis we defer to $\S\ref{s:mcn}$. This is especially
evident from the line width, which for the \NtwoH\ line is $\sim$50\% of
that of the \mcn\ transitions (see Table \ref{tbl:mcnsp}).

\subsubsection{G\,28.87$+$0.06}
\label{ss:totmaps-gte}

The \co\ spectrum of \gte\ strongly suggests the presence of 
high-velocity outflowing gas ($\S$\ref{ss:sp}). 
However, the maps of the wing emission presented in Figure
\ref{fig:gte-maps}a do not seem to outline a clear bipolar structure with
respect to the 3\,mm source position.  A prominent red lobe lies to the
southwest, whereas the blue lobe shows an elongated structure aligned
approximately east-west as well as 
a condensation to the northeast of the main core.  
If we assume that all the CO lobe is associated with an outflow powered by
a source in the core,
the observed morphology and velocity structure can be interpreted in two ways:
(i)~one outflow, with patchy emission from the lobes and with a
large opening angle (close to 90$^\circ$) around the axis denoted by the solid
line in Figure \ref{fig:gte-maps}a;
(ii)~two outflows, well collimated along
the axes represented by the dashed lines in the same figure.
On the basis of
the present data it is impossible to discriminate between these two hypotheses.
Given the existence of several sources in \gte\ ($\S$\ref{ss:contmaps})
and the fact that multiple outflows appear to be frequent in high-mass star forming regions 
(see Beuther et al. 2002b, 2003, 2004),
one cannot rule out the possibility that the blueshifted
condensation to the northeast and some portion of the eastern blueshifted lobe 
are associated respectively with \gte B (and/or C) and D.
However, for the sake of simplicity, in the following we will consider only
the two scenarios proposed above, which assume that it is the main \gte\ core
to host the source(s) powering the outflow(s).\par

In \gte, the peak positions of all tracers but \NtwoH\ roughly coincide with 
the 3\,mm continuum peak (Figure \ref{fig:gte-maps}d).
Both the \amm\ and \mcn\ emission are compact, whereas the \HCO\ emission is
slightly elongated in the north-south direction (Figure \ref{fig:gte-maps}b).
We believe that the \HCO\ emission mostly arises from the HMC and not from the
outflow lobes because its elongation is not parallel to any of the possible
outflow axes and only very faint \HCO\ line wings are detected 
(see Figure \ref{fig:sp}b).  
As done for \gs, we estimate $N_{\rm HCO^+}$ for the region enclosed by the $5\sigma$
level contour assuming two extreme values for the excitation temperature.  We
obtain $N_{\rm HCO^+} =$ $3.9\times 10^{14}$ \cmq\ for \Tex\ $=$ 96\,K, and
$9.2\times 10^{13}$ \cmq\ for \Tex\ $=$ 17\,K, yielding a mass range
\Mlte$\simeq$\,84 -- 360 \Msun\ (Table \ref{tbl:clumps}).  
Here, the maximum \Tex\ is assumed equal to the value of
\Trot\ obtained from the \mcn\ emission.
On the other hand, the minimum \Tex\ is taken equal to the peak \Tsb\ 
of the \HCO\ spectrum 
[14.5\,K from Figure \ref{fig:sp}; 
\Tex $\simeq T_{\rm sb}^{\rm peak}$(\HCO) $+$ \Tbg\ $=$ 17\,K,
where \Tbg\ is the cosmic background temperature].
Independently of the value chosen, this estimate represents
a lower limit because of the different UV coverages,
beam dilutions, and optical depth effects.
We stress that the estimated $N_{\rm HCO^+}$ varies only by a factor of
4.3 regardless of the large uncertainty on \Tex.\par

It is interesting that the \NtwoH\ emission in \gte\ cannot be seen towards
the center of the main core, where instead \amm\ emission is detected (CTC97).  
The distribution of the \NtwoH\ emission is very patchy and several condensations are
seen close to the edge of the \co\ outflow lobes (Figure \ref{fig:gte-maps}c).  
The eastern \NtwoH\ condensation seems to be host \gte D,
although the \NtwoH\ peak position does not agree with the mm peaks.
In fact, none of the \NtwoH\ condensations matches the 3\,mm source positions,
which may indicate a low \NtwoH\
abundances in these cores, as suggested by Womack et al. (1992), who
argue that \NtwoH\ is likely to be destroyed in regions of very high density
($n_{\rm H_2} > 10^6$ \cmc) and high temperature (\Tk $>$ 200\,K).
Alternatively, one might speculate that \NtwoH\ could trace the borders of a region evacuated
by the outflow.  However, our fit to the \NtwoH\ lines give LSR velocities
of around 103 \kms\ for these condensations, namely too similar to the
\Vsys\ to be typical of gas participating in the outflow motion.

\subsubsection{G\,23.01$-$0.41}
\label{ss:totmaps-gtt}

\begin{figure}
\begin{center}
\includegraphics[width=.17\linewidth,angle=0]{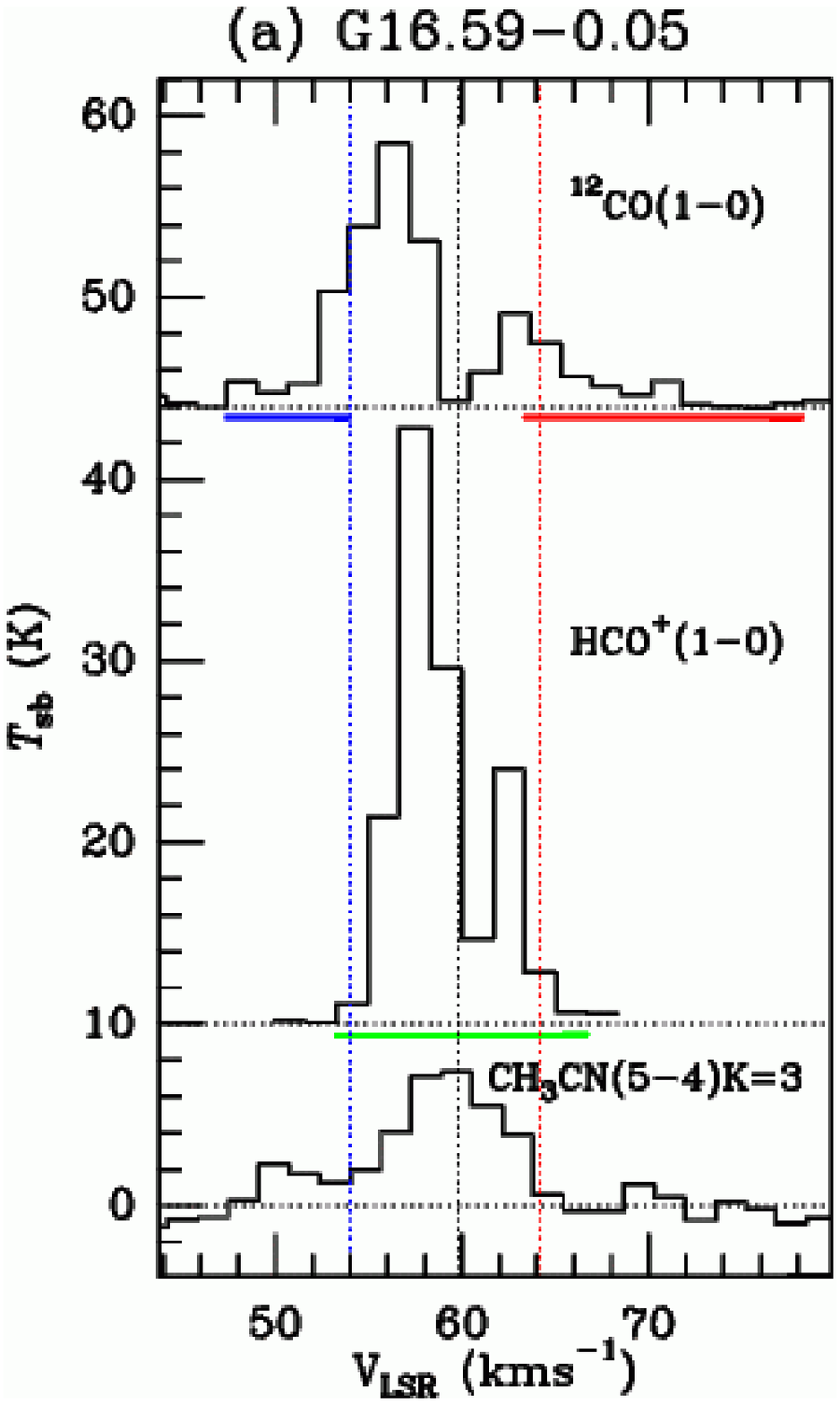} 
\includegraphics[width=.18\linewidth,angle=0]{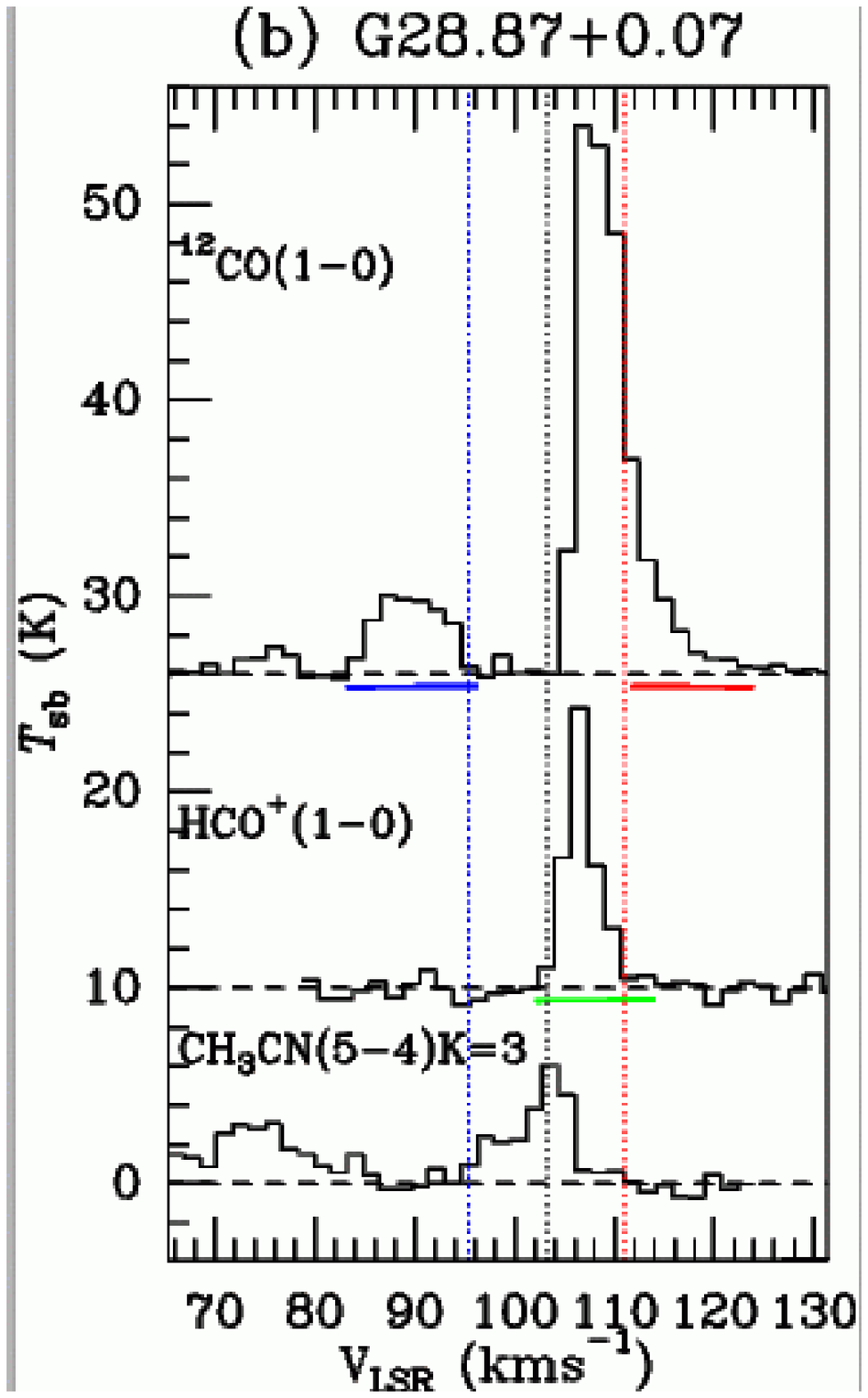} 
\includegraphics[width=.26\linewidth,angle=0]{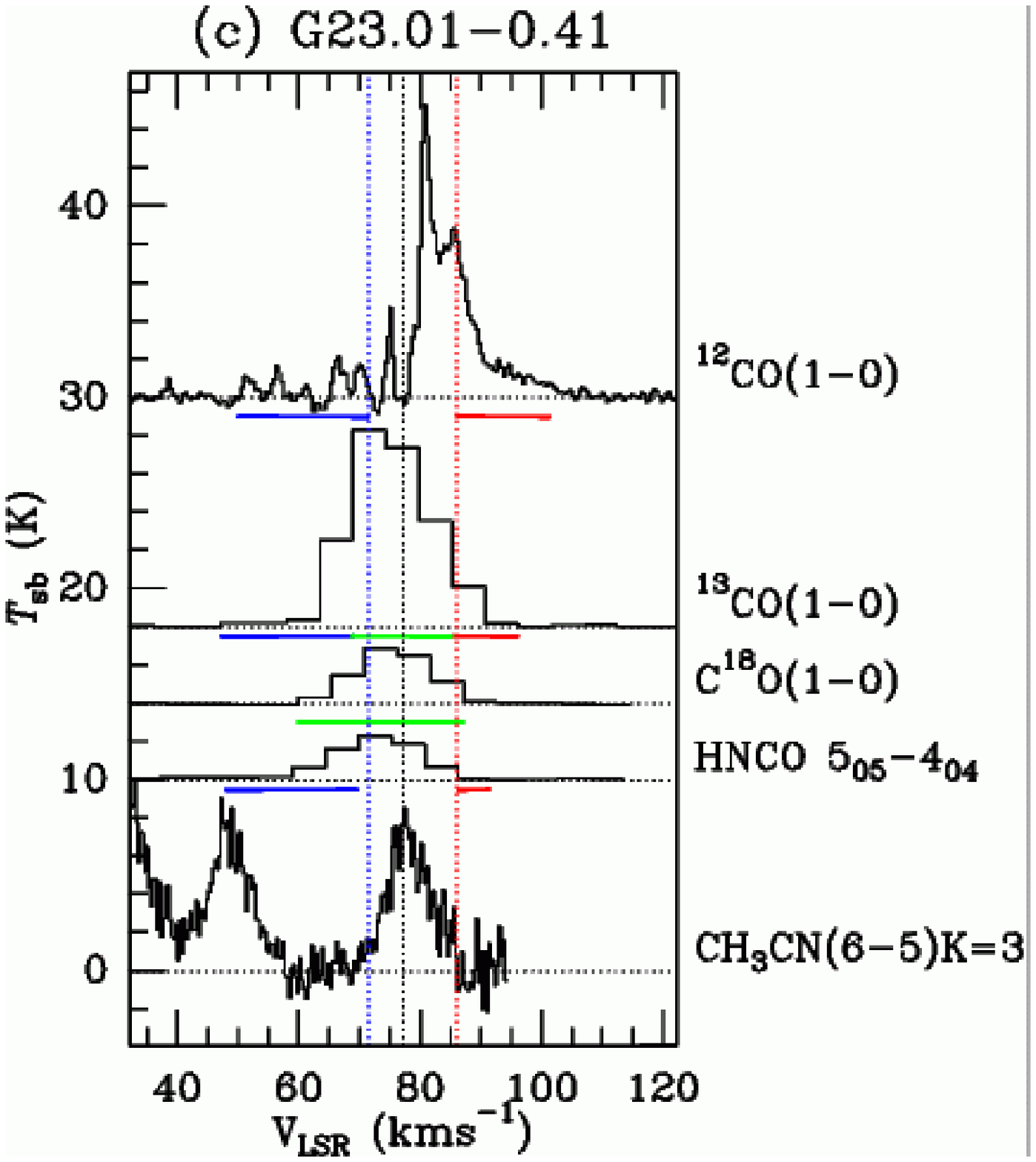} 
\end{center}
\caption{\small Interferometric molecular line spectra towards the 
peak positions of each emission in \Tsb\ scale.
The black vertical dashed lines indicate the \Vsys\ obtained
from the \mcn\ line analysis (Table \ref{tbl:mcnsp}):
\Vlsr\ $=$ 59.8 \kms\ for \gs, 103.3 \kms\ for \gte, and 77.3 \kms\ for \gtt. 
The vertical blue and red dashed lines throughout the panels
show the boundary velocities (\Vb; see $\S\ref{ss:totmaps-gs}$ for the definitions) 
between the core gas and the blue- or redshifted outflow lobes, respectively.
The blue, green, and red bars under the spectra indicate velocity ranges to 
obtain integrated intensity maps 
shown in Figures \ref{fig:gs-maps} -- \ref{fig:gtt-maps}.
The \mcn\ emission seen in \Vlsr\ $\lesssim$ 60 \kms\ at \gtt\ is the other 
$K$-components (see Figure \ref{fig:mcp_sp_integ}).
}
\label{fig:sp}
\end{figure}

\begin{figure}
\begin{center}
\includegraphics[width=.70\linewidth,angle=0]{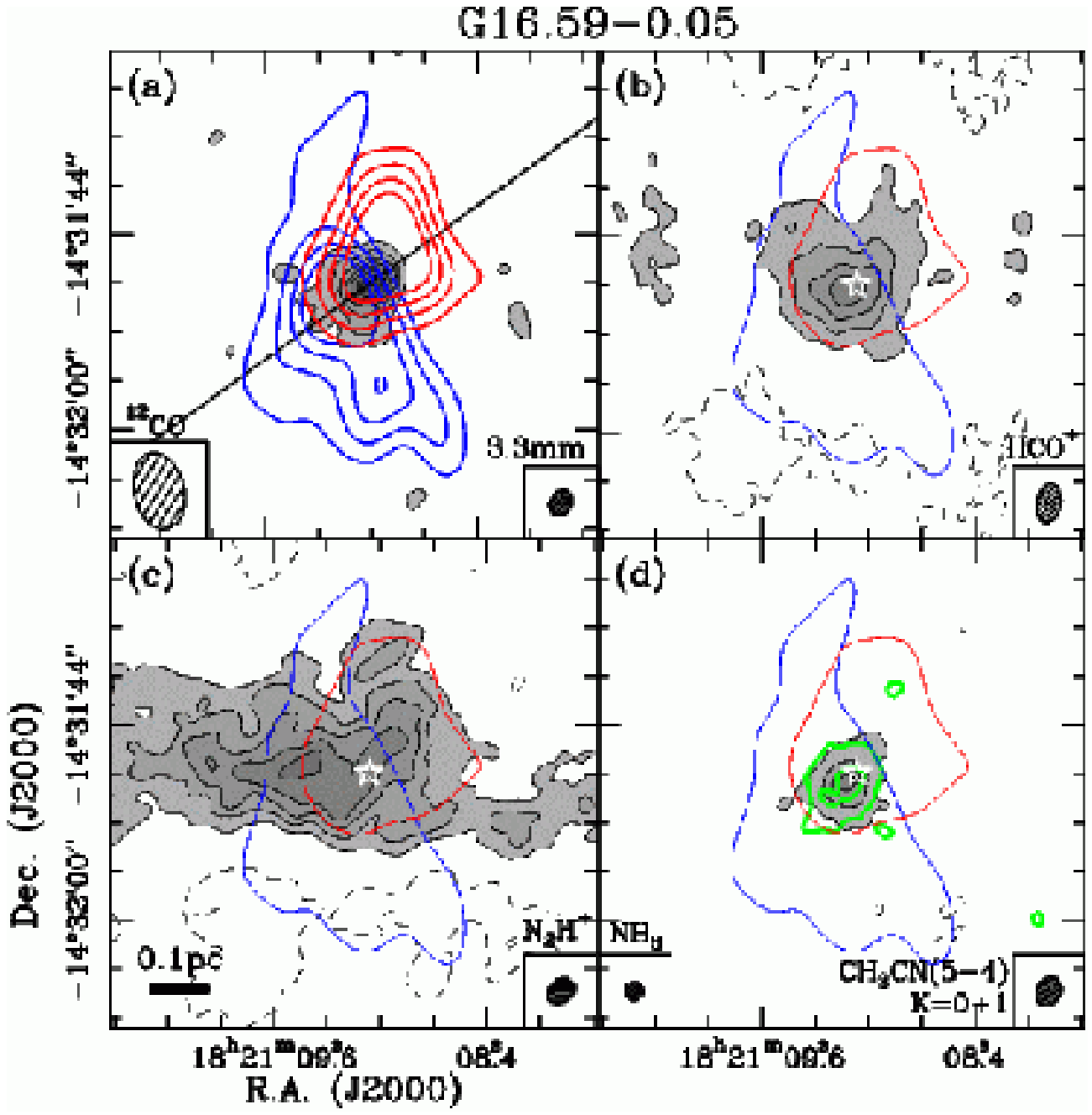}
\end{center}
\epsscale{.80}
\caption{\small Maps of molecular lines and continuum emission for \gs.
(a) Overlays of the blue- and redshifted wing emission maps of
\co\ (1--0) on the 3\,mm continuum emission (grey scale), 
total integrated intensity maps of (b) \HCO\ (1--0) and (c) \NtwoH\ (1--0)
emission (grey scale plus contour), 
and (d) overlay of total maps of \amm\ (3,3) (green contours; CTC97) 
on \mcn\ $K =$ 0 and 1 line emission (grey scale).
In the panels (b)-(d), the thin blue- and red contours indicate the 5$\sigma$
levels of the blue- and redshifted \co\ (1--0) lobes shown in (a).
All the contours, except for the \co, 
start from the 3$\sigma$ level with a 3$\sigma$ step. 
The dashed contours are correspond to $-3\sigma$ and $-6\sigma$ levels
for \HCO\ in panel (b), \NtwoH\ in (c), and \mcn\ in (d). Contours for the
\co\ outflow maps are drawn with 5$\sigma$, 10\sgm, 15\sgm, 20\sgm, and
40\sgm\ levels for the clarity of the plots.  The RMS noise levels of the
images are 33, 19, 0.50, 0.43, 0.10, 4.2, and 0.26 mJy \pbeam\ for
\co\ blueshifted, \co\ redshifted, 3\,mm continuum, \HCO, \NtwoH, \amm, and
\mcn\ maps, respectively.  The solid lines in (a) indicate the identified
outflow axes ($\S\ref{ss:totmaps-gs}$).  The central star in (b)-(d) marks
the peak position of the 3\,mm continuum emission.  The blue and redshifted
\co\ (1--0) emission is integrated over the velocity ranges of 47.4 $\leq$
\Vlsr/\kms\ $\leq$ 54.0 and 63.4 $\leq$ \Vlsr/\kms\ $\leq$ 78.1,
respectively.  The \HCO\ (1--0) line is integrated over 53.3 $\leq$
\Vlsr/\kms\ $\leq$ 66.7.  These velocity ranges are shown by the horizontal
color bars under the spectra shown in Figure \ref{fig:sp}a.  For the
\NtwoH\ (1--0) emission, all the hyperfine emission detected in 49.4 $\leq$
\Vlsr/\kms\ $\leq$ 70.6 is integrated.  The frequency range to obtain the
\mcn\ $K=0+1$ map is shown by a horizontal bar under the corresponding
spectrum in Figure \ref{fig:mcp_sp_integ}.  The ellipse at the lower corners
in each panel shows the synthesized beam size (Tables \ref {tbl:obs_cont} and
\ref{tbl:obs_lin}).  } \label{fig:gs-maps} \end{figure}

\begin{figure}
\begin{center}
\includegraphics[width=.70\linewidth,angle=0]{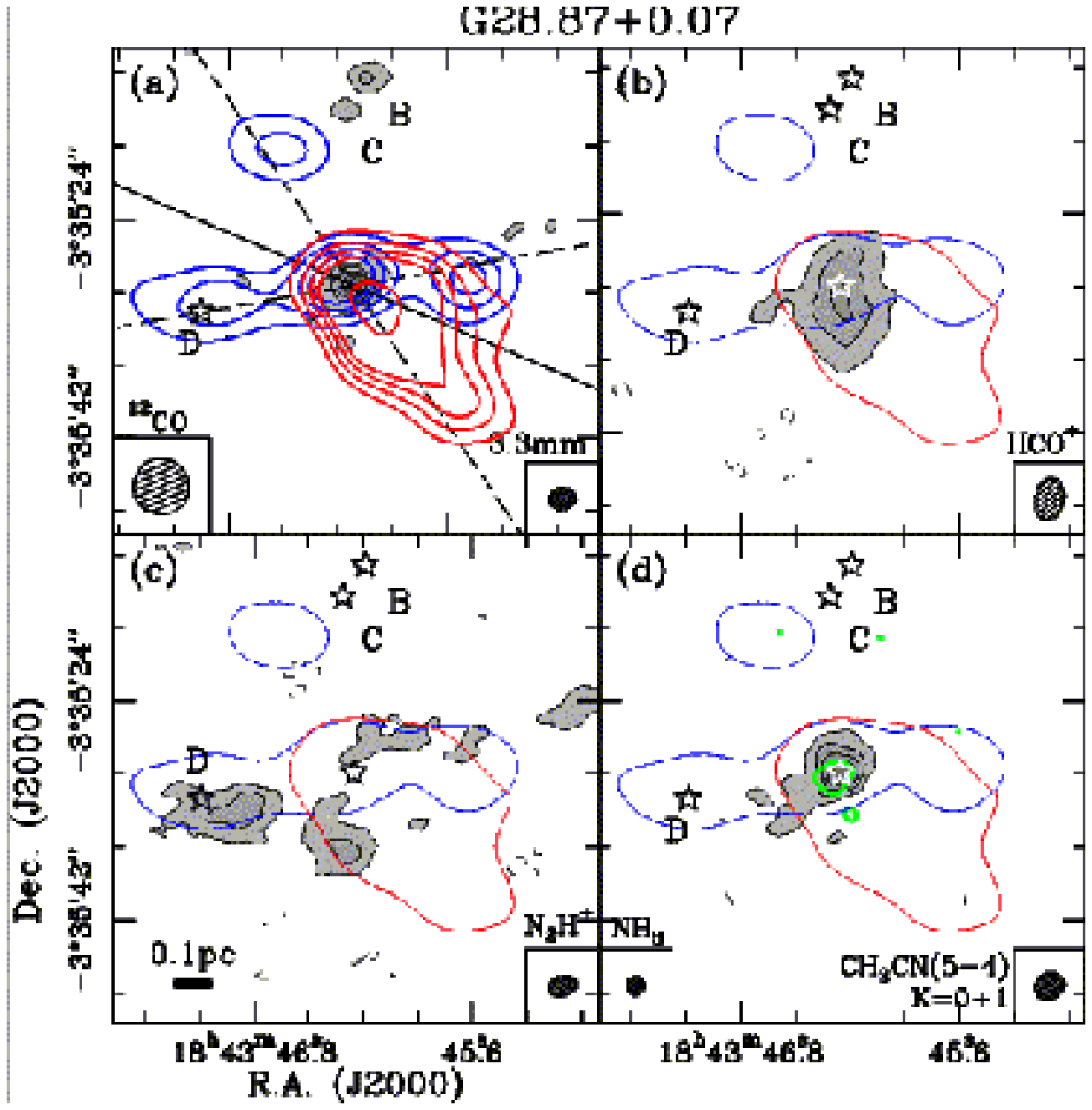}
\end{center}
\epsscale{.80}
\caption{\small Maps of molecular lines and continuum emission for \gte.
All the contours and symbols are the same as those in Figure \ref{fig:gs-maps}.
The RMS noise levels of the images are 
47, 24, 0.51, 0.3, 0.13, 5.0, and 0.18
for \co\ blueshifted, \co\ redshifted, 3\,mm continuum, 
\HCO, \NtwoH, \amm\ (CTC97), and \mcn\ maps, respectively.
The dashed contours correspond to the $-3\sigma$ level
for \HCO\ in panel (b), \NtwoH\ in (c), and \mcn\ in (d).
The blue and redshifted \co\ (1--0) emission, 
\HCO, and \NtwoH\ are integrated over the velocity ranges of 
83.3 $\leq$ \Vlsr/\kms\ $\leq$ 96.3,
111.8 $\leq$ \Vlsr/\kms\ $\leq$ 123.9, 
102.3 $\leq$ \Vlsr/\kms\ $\leq$ 114.0, and
95.0 $\leq$ \Vlsr/\kms\ $\leq$ 111.1, respectively
(see also Figure \ref{fig:sp}b).
The star with the associated labels indicate the peak positions and 
names of the mm sources identified by us (see $\S\ref{ss:contmaps}$ 
and Table \ref{tbl:res_cont}).
All the other symbols are the same as in Figure \ref{fig:gs-maps}.
}
\label{fig:gte-maps}
\end{figure}

\begin{figure}
\begin{center}
\includegraphics[width=.70\linewidth,angle=0]{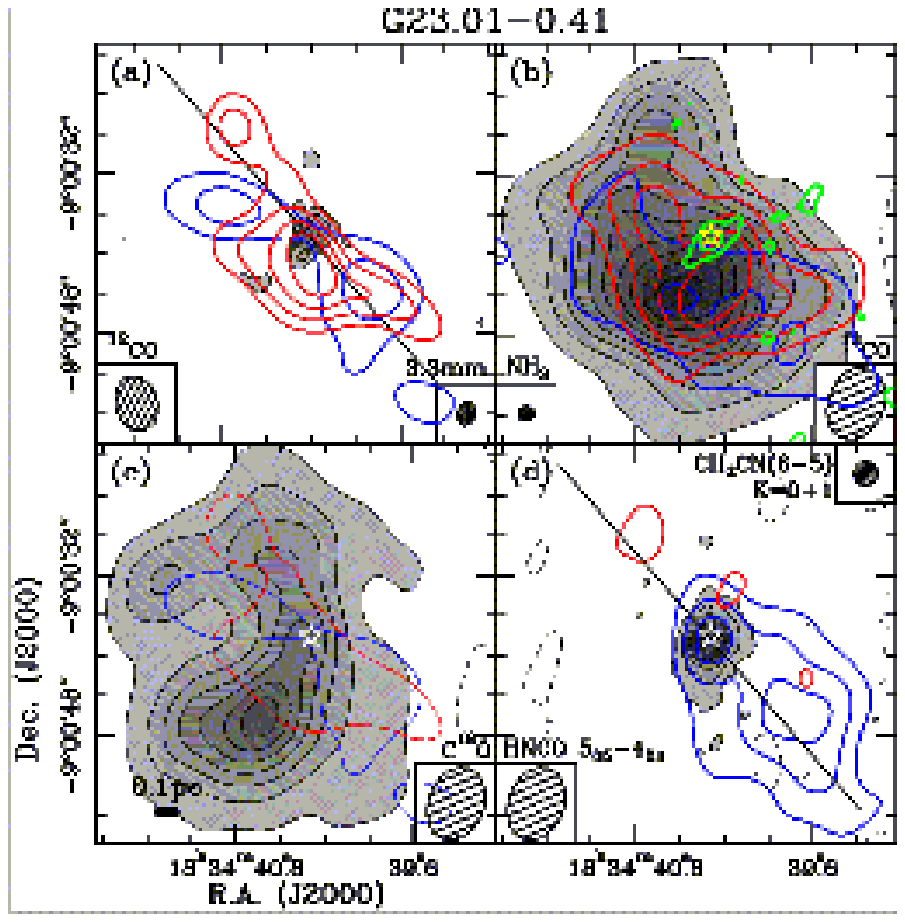}
\end{center}
\epsscale{.80}
\caption{\small Maps of molecular lines and continuum emission for \gtt :
(a) overlay of blue- and redshifted wing emission maps of \co\ (1--0) 
on the 3\,mm continuum emission (grey scale), 
(b) overlay of blue and redshifted wing emission maps of the \tCO\ (1--0), 
total integrated intensity maps of 
\amm\ (2,2) (green contour; CTC97) on the integrated intensity map of the bulk 
\tCO\ (1--0) emission 
(grey scale plus contour; 69.0 $\leq$ \Vlsr/\kms\ $\leq$ 85.3),
(c) total integrated intensity map of \CeO\ (1--0) 
(grey scale; 60.0 $\leq$ \Vlsr/\kms\ $\leq$ 87.2), and 
(d) overlays of the blue and redshifted HNCO $5_{05}-4_{04}$ emission
on the total map of \mcn\ (6--5) $K=0+1$ emission (grey scale).
The dashed contours correspond to the $-3\sigma$ level
for \tCO\ in panel (b), \CeO\ in (c), and HNCO in (d).
The blue- and redshifted wing emission are integrated over the velocity ranges of
49.9 $\leq$ \Vlsr/\kms\ $\leq$ 71.5 and
86.0 $\leq$ \Vlsr/\kms\ $\leq$ 101.6 for \co, 
47.2 $\leq$ \Vlsr/\kms\ $\leq$ 69.0 and
85.3 $\leq$ \Vlsr/\kms\ $\leq$ 96.2 for \tCO, 
48.1 $\leq$ \Vlsr/\kms\ $\leq$ 69.9 and
86.2 $\leq$ \Vlsr/\kms\ $\leq$ 91.6 for HNCO, respectively
(see also Figure \ref{fig:sp}c).
All the contours and symbols are the same as in Figure \ref{fig:gs-maps}.
The RMS noise levels of the images are
90, 26, 1.4, 0.70, 0.79, 5.1, 21, 22 and 0.13 mJy \pbeam\ for 
\co\ blueshifted, 
\co\ redshifted, 
3\,mm continuum, \tCO, \CeO, \amm, 
HNCO blueshifted, HNCO redshifted, and \mcn\ maps, respectively.
All the other symbols are the same as in Figure \ref{fig:gs-maps}.
}
\label{fig:gtt-maps}
\end{figure}

We have seen that the red wing of the \co\ spectrum towards \gtt\ is more
prominent than the blue wing (Figure \ref{fig:sp}c). 
Figure \ref{fig:gtt-maps}a presents an overlay of
the high velocity \co\ emission map on the 3\,mm continuum image;
Figures \ref{fig:gtt-maps}b and d, respectively, show such overlay maps
for the \tCO\ and HNCO wing emissions.
To obtain these maps,
we integrated high velocity emission between the \Vb\ and \Vt\ for each wing
(see $\S\ref{ss:sp}$ for their definitions).
In order to keep a consistency in selecting velocity ranges for the 3 lines
as much as possible,
we integrated the blueshifted \co\ emission emission up to \Vlsr\ $=$ 49.9 \kms\
beyond the \Vt (blue) of 64.6 \kms\ which is the first
LSR-velocity where the blueshifted wing drops below the 1.5\sgm\ level
(see Figure \ref{fig:sp}c).
The resultant velocity ranges for the blueshifted gas are almost
the same for the 3 tracers, while velocity interval for the \co\ red wing is approximately
twice of those for the redshifted \tCO\ and HNCO emission.\par

In Figure \ref{fig:gtt-maps}a, a clear bipolar structure is seen with
the bright redshifted lobe lying to the southwest and the blueshifted lobes to
the northeast and southwest. In addition, a faint redshifted emission is found
to the northeast of the 3\,mm continuum peak.
The velocity structure traced by the \tCO\ and HNCO seems to reconcile with 
that by the \co\ line in the sense that the bright blueshifted gas lies to the
southwest of the 3\,mm source, although the HNCO does not show blueshifted
gas to the northwest. We also point out that 
a fainter redshifted HNCO condensation is seen to the northeast, whose
position matches the northeastern \co\ redshifted lobe.
In addition, one may notice that the redshifted gas outlined by the \tCO\ are basically
same as the \co; both the emission show their peaks to the
south of the 3\,mm source and elongated towards southwest and northeast.\par

As in the case of \gte\ (see $\S\ref{ss:totmaps-gte}$), the above
picture of the high velocity wings may be interpreted either with the
existence of two distinct outflows or with the fact that the outflow axis is
sufficiently close to the plane of the sky to allow the observation of both
blue- and redshifted gas in each lobe.
In the former hypothesis, one of the two putative (proto)stars drives the 
northeastern blueshifted lobe as well as the southwestern redshifted one, 
and the other (proto)star does the remaining two lobes.
If this is the case, each (proto)star is driving a pair of outflow lobes
whose masses (hence, momentum rates, see Table \ref{tbl:outflow})
differs significantly, which seems not to reasonable.
In conclusion, we prefer the single outflow hypothesis with the following 
two supporting reasons:
(1) the high velocity gas maps outlined by the \tCO\ and HNCO 
do not contradict with the velocity structure traced by the \co;
neither \tCO\ nor HNCO maps supports the presence of two outflows, and
(2) we identified only a 3\,mm continuum source towards the center of the HMC
($\S\ref{ss:contmaps}$).
We, thus, adopt the solid line in Figure \ref{fig:gtt-maps}a as the outflow axis.
This line passes across the 3\,mm peak position and as close as possible
to the peak positions of the blue- and redshifted HNCO lobes.\par

In Figures \ref{fig:gtt-maps}b and c we show maps of the \tCO\ and
\CeO\ bulk emission integrated over a velocity range shown by the green bars 
in Figure \ref{fig:sp}c.
These arise from a region ten times
larger than the HMC traced by higher density tracers such as \mcn. 
In addition, the emission appears to peak a few arcsecond to the south of the HMC, which may be
explained if the \tCO\ and \CeO\ lines are partially thick, thus hiding the
densest part of the molecular cloud. To confirm this hypothesis, we have
estimated the optical depth of the \tCO\ and \CeO\ gas from the ratio of the
two isotopomers, assuming a relative abundance of 5.5. Note that our estimate
is not affected by uncertainties due to relative calibration errors or
different UV sampling, because the two lines were observed simultaneously
($\S\ref{ss:ovro}$).
Although the opacity estimate turns out to vary slightly across the line,
depending on the velocity (from $<1$ to 3.4), we conclude that a mean value
of $\tau[$\tCO$ ~(1-0)]\,=\,2$ is appropriate for our purposes. 
This indicates that the optical depth of the \tCO\ (1--0) line may be sufficiently 
high to prevent the detection of the HMC.  
On the other hand, \CeO\ should be less affected by this problem.
In fact a tail of \CeO\ (1--0) bulk emission is seen in Figure \ref{fig:gtt-maps}c 
towards the position of the HMC, although the main peak is still shifted to the south.
This may be also the effect of low angular
resolution, which privileges regions with larger {\it beam averaged} column
density with respect to those (like the HMC) with enhanced volume and {\it
source averaged} column density.\par

We thus obtain mean column densities of 4.2$\times 10^{17}$ \cmq\ for
\tCO\ and 7.1$\times 10^{17}$ \cmq\ for \CeO, over the regions enclosed by
the corresponding 5$\sigma$ contour level, which imply \Mlte\ of 1$\times
10^4$ \Msun\ for \tCO\ and 2$\times 10^4$ \Msun\ for \CeO\ 
(see Table \ref{tbl:clumps}).  
Here we assume that the same \Tex\ for both \tCO\ and \CeO\ and take it
equal to the peak \Tsb\ of the \tCO\ emission 
[ 10.3\,K from Figure \ref{fig:sp}; 
\Tex $\simeq T_{\rm sb}^{\rm peak}$(\tCO) $+$ \Tbg\ $=$ 13\,K].
Despite the lack of zero-spacing information in these interferometric maps,
masses like these compare well to those measured in
similar objects with single-dish telescopes
(e.g., G\,24.78$+$0.08; Cesaroni et al. 2003).\par

Lastly, in Figure \ref{fig:gtt-maps}d we show the map of the \mcn\ emission
integrated over the blue and red bars in Figure \ref{fig:sp}c. 
One may see that the shape of the HMC traced by
this and the \amm\ line (Figure \ref{fig:gtt-maps}b), as well as the 3\,mm
continuum (Figure \ref{fig:gtt-maps}a) is slightly elongated perpendicularly
to the outflow axis. It is thus tempting to speculate that one is observing
a disk-like structure rotating about that axis. We will discuss this issue in
$\S\ref{ss:VelSt}$.

\subsection{Physical Properties of the Molecular Outflows}
\label{ss:outflow}

In order to calculate the outflow properties, we used the \co\ wing emission
maps shown in Figures \ref{fig:gs-maps}--\ref{fig:gtt-maps}.  For \gte, where
multiple outflows might be present (see \S\ref{ss:totmaps-gte}), we consider
only the most prominent lobes, as these represent the dominant contribution
to the mechanical luminosity and momentum rate of the outflow system.  A
caveat to this approach is that the secondary outflow might be weak because
heavily resolved in our interferometric maps. In other words, one cannot rule
out the possibility that better UV sampling at short spacing could reveal a
stronger, better defined outflow pattern. Only single-dish mapping of these
regions may shed light on this issue.\par

The outflow parameters are calculated as follows.  We integrated the outflow
lobe emission encompassing the $5\sigma$ level contours to calculate the
mass of molecular hydrogen constituting the lobes (\Mlobe), 
assuming that the wing emission is optically thin.
Indeed, we have estimated the optical depth in the \co\ line wings 
for \gtt, for which \tCO\ data are available.
We first reconstructed the NMA \tCO\ data with the same CLEAN beam as the PdBI
\co\ data and then smoothed the latter to the same spectral resolution
as the former. From the ratio between the \co\ and \tCO\ line profiles
we computed the optical depth for the blue (47.2 $\leq$ \Vlsr /\kms $\leq$ 69.0)
and red (79.9 $\leq$ \Vlsr /\kms $\leq$ 96.2) wings, which are respectively 
$0.70 \pm 0.2$ and $1.1 \pm 0.5$. This indicates that the correction
due to optical depth effects for the outflow mass and the relevant parameters
is very small, a factor $\sim$1.5.
Given the similarity in the source properties,
we assumed that the wing emission is optically thin also in \gs\ and \gte\
and used the same approach as for \gtt.\par

To calculate the outflow parameters,
as a lower limit for \Tex\ we adopted the maximum \Tsb\
observed in all low-density tracers (CO and \HCO; see Figure \ref{fig:sp}),
namely \Tex\ $=$ 35\,K for \gs, 31\,K for \gte, and 15\,K for \gtt.
Subsequently we calculated the kinematical properties of the flow (see Table
\ref{tbl:outflow}) such as the dynamical time scale (\td), mass loss rate
($\dot{M}_{\rm flow}$), and momentum rate (\Fco). 
The uncertainty on \Tex\ reflects into a 40--70\% uncertainty on \Mlobe.

One can estimate the outflow velocity from the terminal velocity of the CO
spectra and the systemic velocity as $\Vflow=|V_{\rm t}-V_{\rm sys}|$.  The
dynamical time scale is then given by \Llobe/\Vflow, where the lobe length
\Llobe\ is defined as the maximum extent of the lobe measured from the 3\,mm
continuum peak.  In all cases \td\ lies in the range (1--3)$\times 10^4$~yr.
Note that this estimate of \td\ is to be taken as a lower limit because the
lobe could extended beyond our field of view. \par

The mass loss rate, $\dot{M}_{\rm flow}$, was estimated from the ratio
$M_{\rm lobe}$/\td, and turns out to be on the order of
$10^{-4}$ -- $10^{-3}$\Msun\ yr$^{-1}$ for all sources.
Since molecular outflows appear to be momentum-driven (e.g., Cabrit \&
Bertout 1992) the momentum rate \Fco=\Mlobe$V_{\rm flow}^2$/\Llobe\
may be taken as an indicator of the strength of the outflow and hence
of the mass and luminosity of the YSO powering it.
The \gs\ outflow has $\Fco\simeq10^{-3}$ \Msun\ \kms\ yr$^{-1}$,
whereas for \gte\ and \gtt\ \Fco\ are an order of magnitude higher. Values like these
are typical of YSOs with luminosities of $\sim 10^3-10^4$ \Lsun\
(see Figure~5 of Richer et al. 2000 and Figure~4 of Beuther et al. 2002a),
confirming that we are indeed dealing with early B (proto)stars.\par

These findings are not affected if one takes into account the unknown
inclination angle $i$, here defined as the angle between the outflow axis and
the line of sight.  In fact, \td, $\dot{M}_{\rm flow}$, and \Fco, are
proportional respectively to $\cot i$, $\tan i$, and $\sin i/\cos ^2i$. Thus,
assuming as extreme values $i=20\degr$ and $i=70\degr$,
the corrections for the three quantities above are respectively 2.7--0.36,
0.36--2.7, and 0.39--8. In conclusion, for all reasonable inclination
angles, the parameters listed in Table~7
might need a correction by less than an order of magnitude, thus
leaving our conclusions unaffected.

\subsection{Temperature, Column Density and \mcn\ Abundance of Cores}
\label{s:mcn}

\subsubsection{Rotational Temperature and Column Density of the \mcn\ Lines}
\label{ss:rd}

Symmetric top molecules like \mcn\ are ideal probes to measure the gas
kinetic temperature, because different $K$ components belonging to the same
$J+1\rightarrow J$ transition are mainly excited by collisions, and
spread over a few 10~MHz and can be
observed simultaneously in the same bandwidth 
(hence, suffering less calibration errors). 
In addition, for densities typical of
HMCs ($>10^{6-7}$~\cmc) the rotational temperature, \Trot, obtained from $K$ line
ratios turns out to be very close to the kinetic temperature of the H$_2$ gas.  
One may thus obtain reliable estimates of the temperature and column
density of the region where \mcn\ is detected through the rotation diagram
method (Hollis 1982; Churchwell, Walmsley \& Wood 1992).\par

Since optical depth effects may affect the reliability of rotation diagrams,
we have attempted to fit the $K$ components of \mcn\ with a simple LTE fit
taking into account also the line opacity, as performed in e.g., Olimi et al. (1993). 
The results indicate that the
optical depth is less than $\sim2$, sufficiently low to allow
usage of the rotation diagram method in our cases.

\begin{figure}
\begin{center}
\includegraphics[width=.35\linewidth,angle=-90]{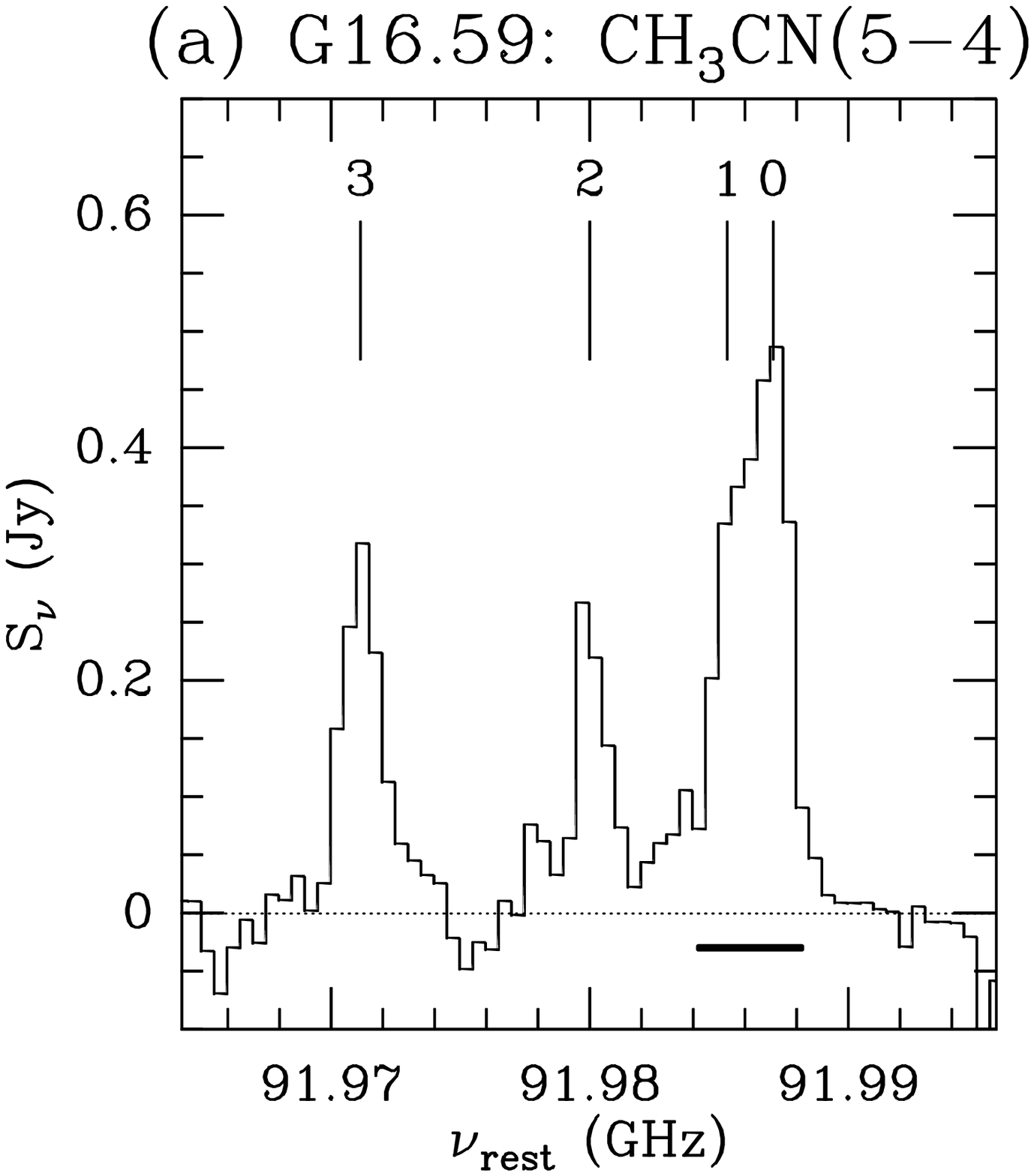} \\
\includegraphics[width=.35\linewidth,angle=-90]{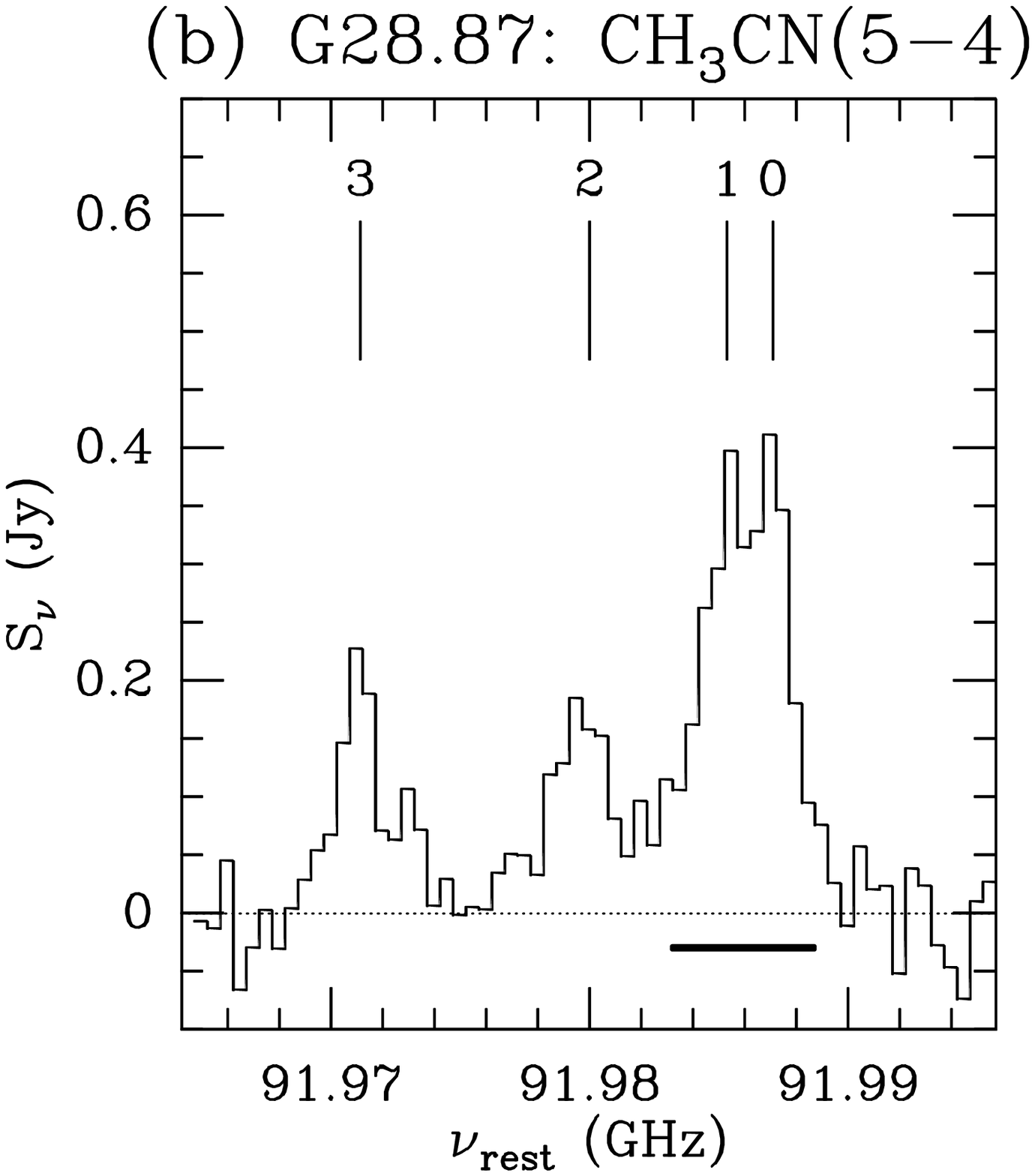} \\
\includegraphics[width=.32\linewidth,angle=0]{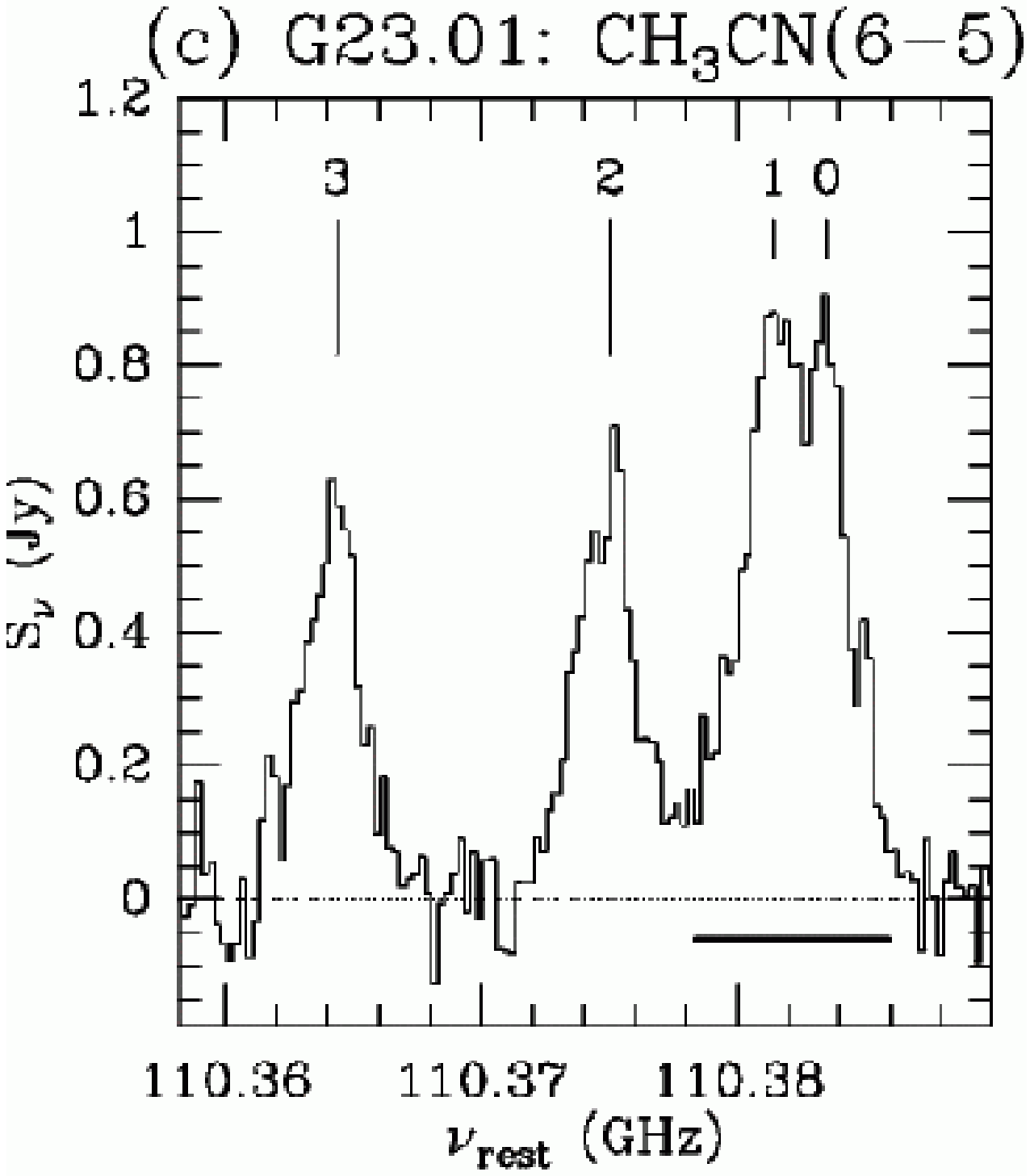} \\
\end{center}
\caption{\small 
Mean \mcn\ spectra integrated over the 5$\sigma$ level areas of the 3 HMCs
in flux density (\Snu) scale:
$J = 5$--4 for G\,16.59$-$0.05 and G\,28.87$+$0.07, and $J = 6$--5 for G\,23.01$-$0.41.
The vertical bars above the spectra indicate the rest-frequency of 
the $K$ rotational ladder emission. 
The horizontal bars under $K=$ 0 and 1 emission indicate frequency ranges to
obtain the integrated intensity maps in the panels (d) of
Figures \ref{fig:gs-maps} -- \ref{fig:gtt-maps}.
}
\label{fig:mcp_sp_integ}
\end{figure}

\begin{figure}
\begin{center}
\includegraphics[width=.41\linewidth,angle=-90]{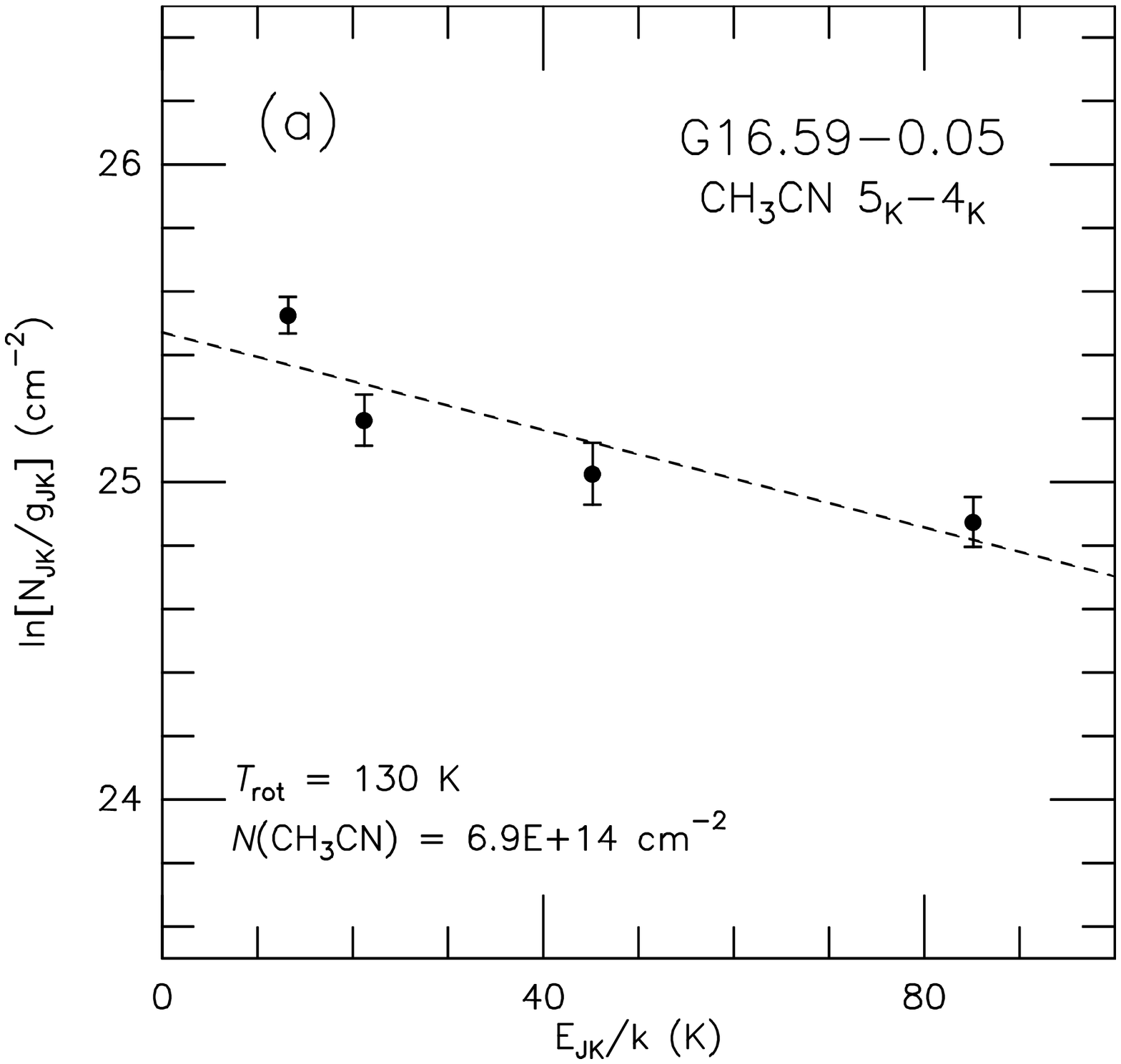}
\includegraphics[width=.41\linewidth,angle=-90]{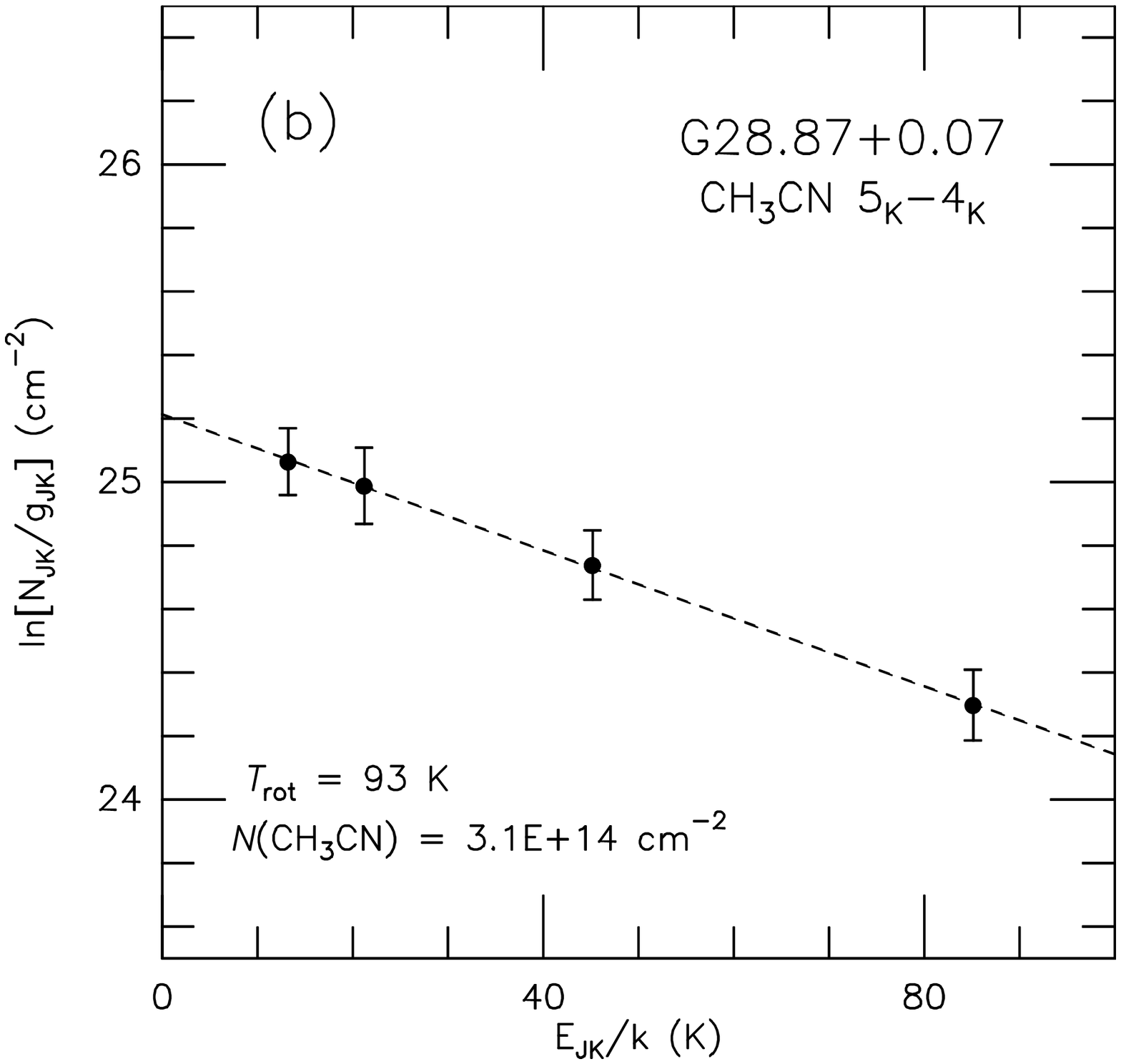}
\includegraphics[width=.41\linewidth,angle=-90]{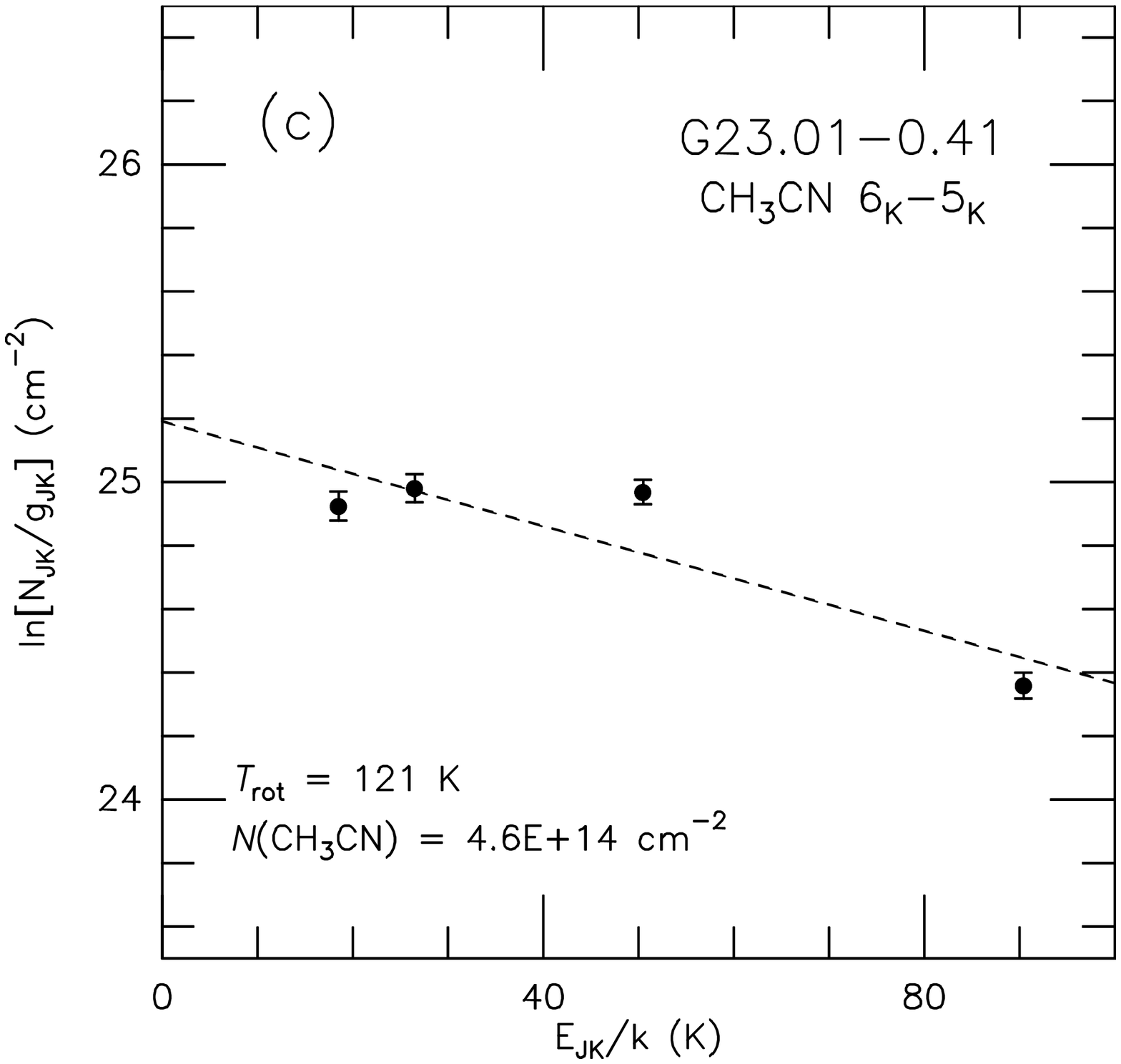}
\end{center}
\epsscale{.80}
\caption{Rotational diagrams for \mcn\ $J_K\rightarrow (J-1)_K$ transitions 
obtained from the averaged spectra over the 5$\sigma$ areas of the $K=0+1$ 
emission (Figure \ref{fig:mcp_sp_integ}), see $\S\ref{ss:rd}$.
The natural logarithm of the column density in state $J$ per substate 
[$\ln (N_{\rm JK}/g_{\rm JK})$] (\cmq)
is plotted to the upper energy level of $E_{\rm J}/k$.
The dashed straight line is a least-square fit to the data, 
yielding rotational temperature (\Trot) and column density, $N(\mcn )$, shown 
in each panel. All the derived parameters are summarized in Table \ref{tbl:res_mcn}.
}
\label{fig:Trot}
\end{figure}

To derive mean values of \Trot\ and \Nmcn\ over the cores using
rotation diagrams, we analyzed the
mean spectra in Figure \ref{fig:mcp_sp_integ} obtained by averaging the
emission over the regions inside 5$\sigma$ contour level of the $K =0+1$
emission map.  Assuming that all $K$-components trace the same gas, we
performed multiple-Gaussian fitting of the lines, 
forcing their widths to be identical and their separations 
in frequency to be equal to the laboratory values. 
Hence, the free parameters of the fit are the line
intensities, the full width at half maximum (FWHM) of the lines, and the LSR
velocity of one $K$ component -- arbitrarily chosen.  
Tables \ref{tbl:mcnsp} and \ref{tbl:res_mcn} summarize the fitting results, 
where \Vsys\ is the
best-fit LSR velocity.  The line intensities were used to make rotation
diagrams (see e.g. Churchwell, Walmsley \& Wood 1992) from which we obtained
an estimate of \Trot\ and \Nmcn.  Note that our analysis is restricted to use
the $K =$ 0 to 3 lines because of the limited correlator bandwidth.  In these
calculations, we assumed optically thin emission and adopted the partition
function by Araya et al. (2005).  The rotation diagrams for all three sources
are presented in Figure \ref{fig:Trot}, where $N_{\rm JK}$, $E_{\rm JK}$, and
$g_{\rm JK}$ represent the column density, energy, and statistical weight of
level $(J,K)$, and $k$ is the Boltzmann constant.  
In Table \ref{tbl:res_mcn}, we give the obtained values of \Trot\ and \Nmcn\ 
for the 3 HMCs.\par

The 3 HMCs have comparable mean \NMCN, spanning the range (3--7)$\times
10^{14}$ \cmq\ (Table \ref{tbl:res_mcn}), close to the values obtained
from interferometric measurements in similar objects 
(e.g. G\,24.78$+$0.08; Beltr\'an et al. 2005).
The mean values of \Trot\ in \gs\ and \gtt\ (120 -- 130\,K) are likely 
to be higher than that of \gte\ (93\,K), although our \Trot\ estimates have
rather large errors.
In \gte\ both the temperature estimated from \amm\ (see Table \ref{tbl:res_mcn}) 
and the \Tsb\ of the \mcn\ lines 
($\sim 5$\,K for $K=0+1$) are twice smaller than in the other two sources
($\sim 10$\,K), suggesting that \mcn\ gas should be colder than in \gs\ and \gtt.\par

In Table \ref{tbl:res_mcn}, one may notice that the mean \Trot\ are a factor
of 2 higher than the \amm\ kinetic temperature (\Tk) derived from the VLA
observations (CTC97).  One possible explanation for this discrepancy is that 
optical depth effects in the \mcn\ lines may cause an overestimate of \Trot. 
It is also worth noting that the low-energy
\amm\ transitions observed by CTC97 may be affected by more extended emission
(and hence colder gas) than those obtained through the \mcn\ lines, as
suggested in Olmi et al. (1993). In fact, these authors
demonstrate that the two ``thermometers'', \mcn\ and \amm, are in reasonable
agreement for $T \lesssim 50$\,K, but for higher temperature the
\Trot\ estimated from \mcn\ tends to exceed that from \amm.  However, one
should keep in mind that the relationship of Olmi et al. (1993) is based on
single-dish observations with 6--8 times larger beam sizes than our
interferometric measurements, and might hence be affected by material
extended over larger regions than those imaged by us.

\subsubsection{Fractional Abundance of \mcn\ Molecules in  the Cores}
\label{ss:Xmcn}

We have attempted an estimate of the fractional abundance of \mcn\ (\Xmcn) from the
ratio between the mean \NMCN\ (Table \ref{tbl:res_mcn}) and
\meanNmol\ calculated in $\S$\ref{ss:sed}.  Table \ref{tbl:res_mcn} gives the
resultant \Xmcn.  We stress that the 3\,mm emission from the three HMCs is
dominated by thermal dust emission ($\S$\ref{ss:sed}).
Note that the \mcn\ line and 3\,mm continuum emission show fairly similar spatial
distributions (see panels (a) and (d) in Figures \ref{fig:vmaps-gs}--\ref{fig:vmaps-gtt}).
It is also worth pointing out that the errors on \Xmcn\ take into account only uncertainties
on the flux calibration ($\S\S\ref{ss:ovro}$ and \ref{ss:nma}).  One can see that \Xmcn\ lies
in the range $(0.7-2)\times 10^{-9}$, marginally lower than the values
quoted for the well known HMC in Orion ($5\times 10^{-9}$; see van Dishoeck et al. 1993).

\subsection{Velocity Structure and Stability of the \mcn\ Cores}
\label{ss:VelSt}

Our results demonstrate that the ammonia cores imaged by CTC97 are indeed
compact ($\simeq 0.1$ pc), dense ($\simeq 10^{6}$ \cmc), and hot ($\simeq 100$ K)
molecular cores.
This strengthens the idea that high-mass (proto)stars -- whose presence is suggested
by the presence of \wat\ and OH masers (Foster \& Caswell 1989) --
are embedded in such massive ($\simeq 100$ \Msun) HMCs.
In the case of \gtt, we also find
evidence of a large scale clump seen in two CO isotopomers enshrouding the HMC, 
which is consistent with the findings of other authors (see e.g. Fontani et al.
2002 and references therein) in similar objects.
In this section we wish to perform a more detailed analysis of the HMC
structure by focusing on the velocity field of the gas, with the purpose of
studying the HMC stability and infer the possible presence of rotation.\par

\begin{figure}
\begin{center}
\includegraphics[width=.42\linewidth,angle=-90]{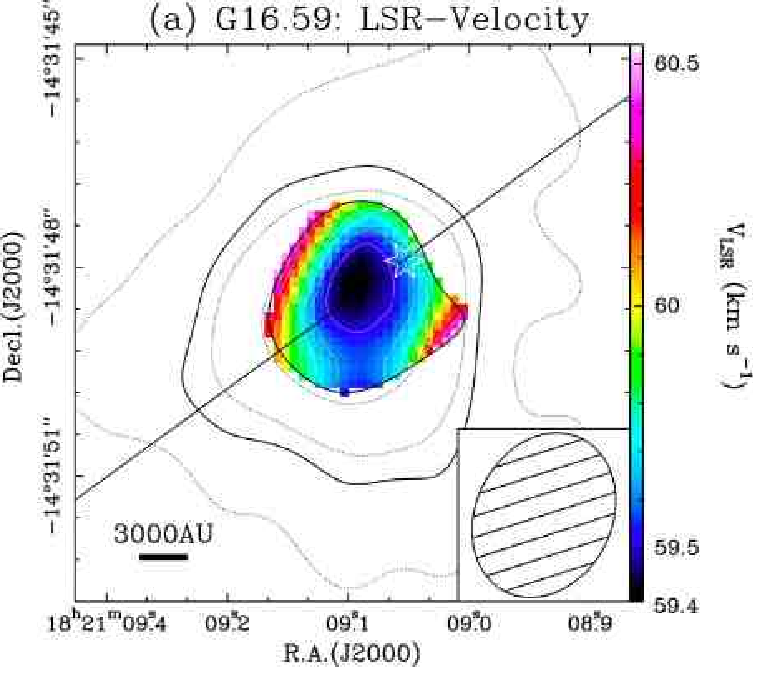}
\includegraphics[width=.42\linewidth,angle=-90]{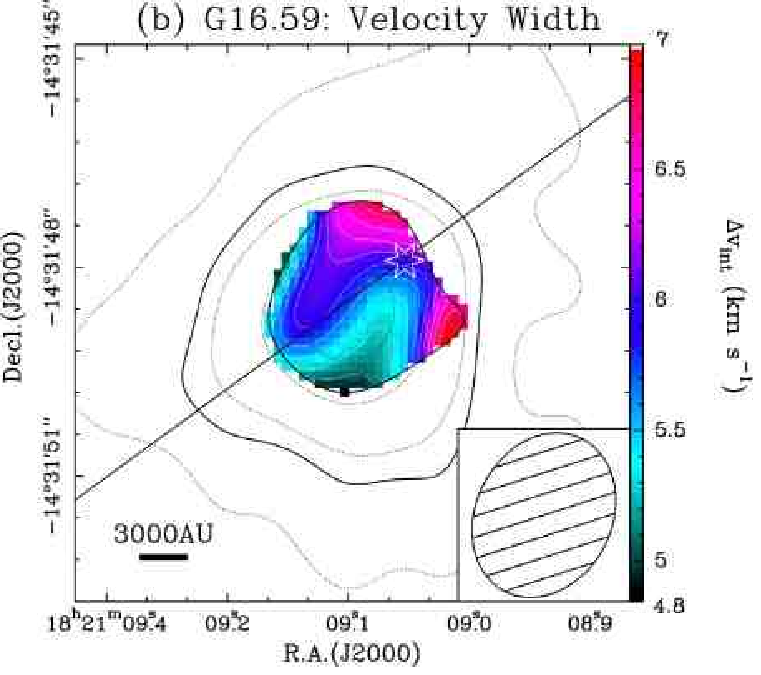}
\end{center}
\epsscale{.80}
\caption{\small 
(a) Isovelocity and (b) intrinsic velocity width (\dVint) maps 
obtained from the multiple Gaussian line-profile fitting to the \mcn\ emission 
towards the central portion of the \gs\ HMC.
These maps are presented inside the 5$\sigma$ level contour 
(inner solid contour) of the $K =$ 3 total maps;
the outer solid contour indicates the $5\sigma$ level of the $K=0+1$ map.
The dashed thin contours show the positive contour for the $K =$ 0$+$1 
map in Figure \ref{fig:gs-maps}.
The ellipses at the bottom-right corners indicate 
synthesized beam size (Table \ref{tbl:obs_cont}).
The straight lines and central stars are the same as in Figure \ref{fig:gs-maps}.
Systemic velocity (\Vsys) of the HMC is \Vlsr = 59.8 \kms\ (Table \ref{tbl:mcnsp}),
and thermal line width at \Tk\ $=$ \Trot\ $=$ 130\,K is 0.38 \kms.
}
\label{fig:vmaps-gs}
\end{figure}

\begin{figure}
\begin{center}
\includegraphics[width=.42\linewidth,angle=-90]{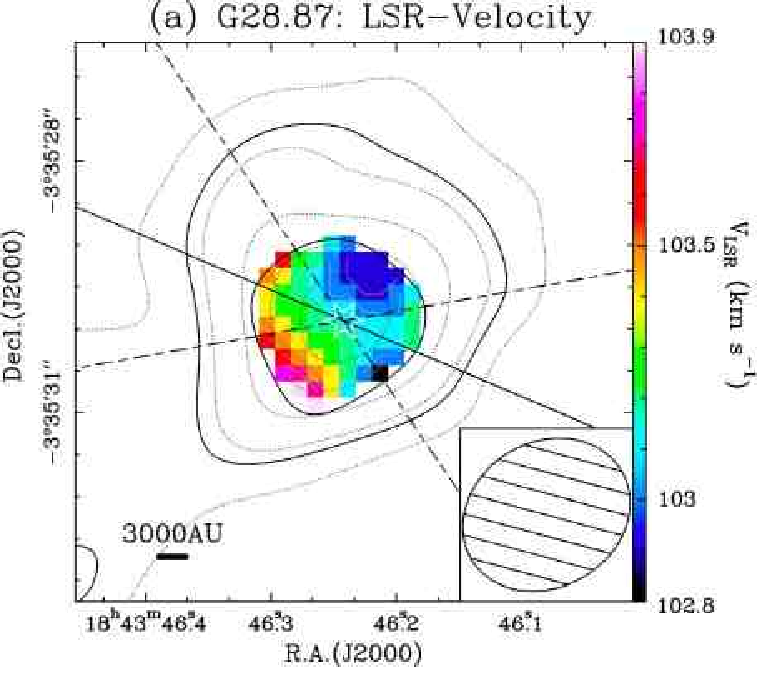}
\includegraphics[width=.42\linewidth,angle=-90]{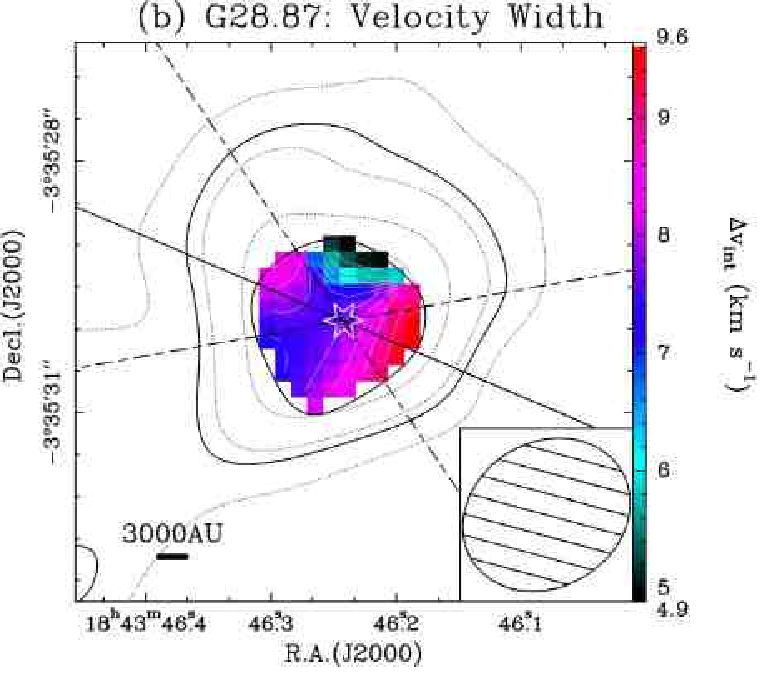}
\end{center}
\epsscale{.80}
\caption{\small Same as Figure \ref{fig:vmaps-gs}, but for \gte.
The \Vsys\ is \Vlsr = 103.5 \kms\ (Table \ref{tbl:mcnsp}), and thermal
line width at 93\,K is 0.32 \kms.
}
\label{fig:vmaps-gte}
\end{figure}

\begin{figure}
\begin{center}
\includegraphics[width=.42\linewidth,angle=-90]{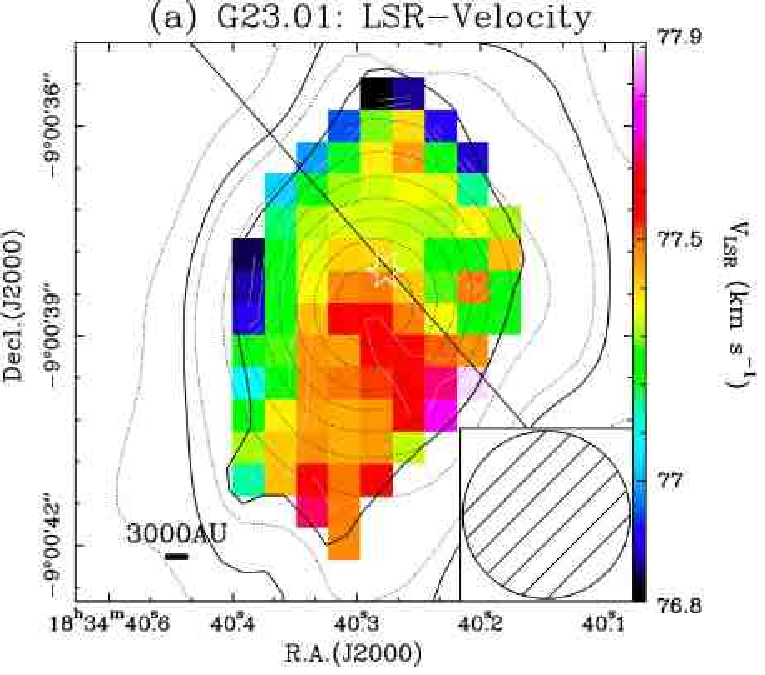}
\includegraphics[width=.42\linewidth,angle=-90]{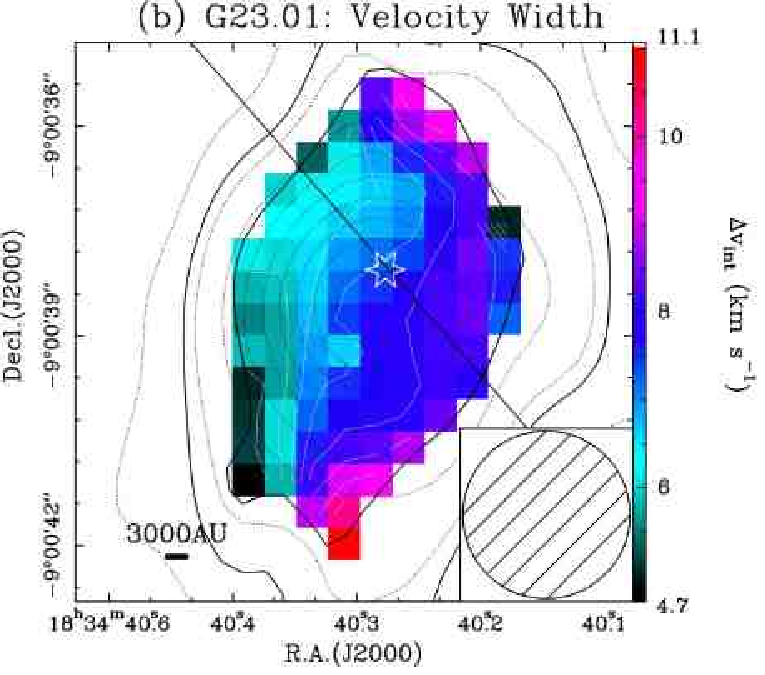}
\end{center}
\epsscale{.80}
\caption{\small Same as Figure \ref{fig:vmaps-gs}, but for \gtt.
The transition employed here is $J=$6--5.
The \Vsys\ is \Vlsr = 77.4 \kms\ (Table \ref{tbl:mcnsp}), and thermal
line width at 121\,K is 0.37 \kms.
}
\label{fig:vmaps-gtt}
\end{figure}

To gain insight into the velocity field of the gas, we used the same approach as
Beltr\'an et al. (2004), i.e. we determined the velocity in each point of the
region where \mcn\ is detected, by fitting the \mcn\ spectrum in such points
with the method described in \S\ref{ss:rd}. In this way we also obtained the
\mcn\ line width (\dVint) in each point of the map: this is the FWHM 
after deconvolution of the instrumental resolution (Table \ref{tbl:obs_lin}). 
In this process, we have considered only those points for which the intensity of the
$K=3$ line was above 5$\sigma$.  The results are shown in Table
\ref{tbl:CoreCenter} and Figures \ref{fig:vmaps-gs}--\ref{fig:vmaps-gtt}.  
It should be also noted that the regions over which the \mcn\ velocity could be determined are only
barely resolved, being comparable to the synthesized beam width.  

\paragraph{G\,16.59$-$0.06} 
The velocity field map of \gs\ (Figure \ref{fig:vmaps-gs}a) shows a gradient
approximately along a line with P.A.$\simeq$ 45\degr, roughly perpendicular
to the outflow axis.  The most blueshifted gas lies at the center, while the
most redshifted is seen to the northeast and southwest, close to the
border of the region. 
A map of the line width is shown in Figure
\ref{fig:vmaps-gs}b and appears to have no correlation with the
velocity map.\par

It would be surprising if the velocity gradient in this core were related
to the CO outflow shown in Figure \ref{fig:gs-maps}, because the two are
inconsistent both geometrically and kinematically. On the other hand, it is
difficult to believe that the \mcn\ velocity is tracing rotation, since in
this case the most blue- and redshifted gas should be found at the opposite
extremes of the gradient. Furthermore, the velocity gradient corresponds to
1.2 \kms\ over 2400 AU, which implies an equilibrium mass of 0.2~\Msun, much
less than the mass of the HMC.\par

We believe, however, that the presence of rotation about the outflow axis
cannot be ruled out on the basis of our observations. In fact, the outflow
appears to be close to pole-on, which would make the component along the 
line-of-sight of the rotation velocity very small and hence difficult to measure.\par

For a better understanding of the origin of the velocity gradient, one should
also explain the positional displacement between the 3\,mm and \mcn\ peaks.
We propose two hypotheses for follow-up studies: 
\begin{itemize}
\item~The continuum and \mcn\ emission might arise from two distinct YSOs,
the former deeply embedded in a relatively cold, massive core, the latter
associated with hot gas, possibly distributed in a rotating ring, similar
to those seen around binary systems in low-mass stars (e.g. GG~Tau; Guilloteau
et al. 1999); the lack of gas and dust at the center of the ring would explain
why the continuum emission from the \mcn\ peak is faint.
\item~If the newly formed stars lie close to the surface of the core, this
might cause a peak in the \mcn\ abundance close the the stars themselves
and far from the core center, where the dust column density -- and hence the
mm continuum emission -- peaks.
In this scenario, the 3\,mm peak
could represent a high-mass protostar, while the \mcn\ peak would correspond
to a relatively more evolved OB star.
\end{itemize}

\paragraph{G\,28.87$+$0.08}

The velocity map of the \gte\ core (Figure \ref{fig:vmaps-gte}a) 
displays a clear gradient from southeast to northwest plus a less
prominent one from northeast to southwest. 
It is worth reminding that the 3\sgm\ contour levels of the 2.6\,mm
continuum (Figure \ref{fig:contmaps}b) and \mcn\ $K=0+1$ emission 
(Figure \ref{fig:gte-maps}d) are slightly elongated in a
direction roughly perpendicular to the outflows
($\S$\ref{ss:totmaps-gte}).
Such an elongation is consistent with a flattened circumstellar toroid.\par

The interpretation of the isovelocity map is complicated 
by the fact that the number of possible outflows and their
structure is unclear in this region (see \S\ref{ss:totmaps-gte}).
However, there are a few points worth of consideration:

\begin{itemize}

\item The velocity gradients seen in \mcn\ are inconsistent with the outflow
 velocity -- no matter whether the one- or two-outflow hypothesis is correct --
 because the redshifted \mcn\ gas is located approximately to the E-SE
 of the core, whereas the redshifted CO emission is found to the W-SW.

\item If a single outflow is present, its axis (solid line in Figure \ref{fig:vmaps-gte}a)
 runs perpendicular to the most prominent \mcn\ gradient.

\item If instead two outflows are present, neither of these would be orthogonal
 to either \mcn\ velocity gradient.

\end{itemize}

Assuming that the \mcn\ emission is tracing a core rotating 
about the axis outlined by a poorly collimated CO outflow,
the single outflow hypothesis in $\S\ref{ss:totmaps-gte}$ 
seems preferable to
the double outflow scenario.
Additional evidence in favor of this scenario is given by the line width
map of Figure \ref{fig:vmaps-gte}b, which shows that significant line
broadening, by $\sim$2\,km\,s$^{-1}$, occurs from the core center to the border,
just along the outflow axis.\par

The \mcn\ velocity gradient corresponds to a change of $\simeq 1.1$
\kms\ over $\simeq 1.1\times 10^4$ AU.  If this is due to
rotation, the dynamical mass ($\Mdyn\ = R_{\rm eff}V_{\rm rot}^2/G$ with
\Vrot\ rotational velocity; Table \ref{tbl:CoreCenter}) required to balance centrifugal 
and gravitational forces is equal to $3\pm 2$ \Msun\ (Table \ref{tbl:CoreCenter}). 
This is an order of magnitude smaller than mass of the core estimated from
3\,mm continuum flux (\Mket\ $\simeq$ 55 \Msun; Table \ref{tbl:CoreCenter}),
over the region where the rotation velocity is measured,
and two orders of magnitude less than the virial mass
($\Mvir\simeq R_{\rm eff}(\dVint)^2/G \simeq 1100$ \Msun; Table \ref{tbl:CoreCenter})
obtained for the same region.
This fact strongly suggests that even though rotation may 
be present in the HMC, its contribution
to gravitational equilibrium is negligible.

What does support the core against gravitational collapse?
Here we should recall that the \dVint\ map of Figure \ref{fig:vmaps-gte}b does 
not show a well-defined increase of the velocity width towards the core center,
as one sees, instead, in similar objects -- e.g., G\,31.41$+$0.31 and G\,24.78$+$0.08
(Beltr\'an et al. 2005; see their Figures 9 and 23).
Therefore, our data do not provide any evidence 
for infall motions towards the core center, so that infall is unlikely
to contribute to line broadening.
A more likely explanation for the observed line width is that of
turbulent motions of the core gas, as
turbulence is believed to play a dominant role on the evolution of the core
and the star formation process in its interiors (e.g., McKee \& Tan 2003).

\paragraph{G\,23.01$-$0.41} 
The \gtt\ core is significantly elongated in the southeast-northwest direction 
(see Figure \ref{fig:gtt-maps}d; P.A.$\simeq$ 160\degr), 
namely perpendicular to the outflow axis, as one can see
in the \amm, \mcn, and 3.3\,mm continuum maps in Figure \ref{fig:gtt-maps}. In
the same figure, one can appreciate that also the \CeO(1--0) line presents a
tail of emission in the same direction, extending towards the position of
the 3\,mm continuum peak. All the results strongly suggest that one could be
observing a flattened structure rotating about the outflow axis. To shed light
on this issue, as already done for the other two sources, we plot in
Figure \ref{fig:vmaps-gtt}a the velocity field of the \mcn\ core. 
Regardless of some scatter, the velocity appears to increase quite steadily along the
major axis of the HMC, as expected in case of rotation about the outflow
axis (denoted by the solid line in the figure). The velocity shift amounts
to 1.1~\kms\ over $2.6\times10^4$ AU.\par

Unlike the velocity, the line width increases from north-east to south-west, 
roughly along the outflow axis. This suggests that to some extent the \mcn\ emitting gas
might also participate in the expansion. This fact could explain why the most
blue- and redshifted \mcn\ emission is observed respectively to the north and
south of the HMC, although the main velocity trend lies along the SE--NW
direction.  We conclude, that, albeit slightly affected by the outflow, the
\mcn\ line emission from the HMC is mostly tracing rotation.\par

As already done for \gte, we have estimated the mass needed
for rotational support of the core. This is $\Mdyn\simeq 8\pm 3$ \Msun,
two orders of magnitude smaller than the HMC mass $\Mket\simeq340$~\Msun\
(Table \ref{tbl:CoreCenter}). Instead, the virial mass is 5 times larger than
the core mass, demonstrating that as well as for \gte,
also for \gtt\ the dominant contribution to core equilibrium is not
coming from rotation but from turbulence.\par

\subsection{Comparison with the Other \mcn\ Cores and Future Considerations}
\label{ss:nature}

As discussed in $\S$\ref{ss:VelSt}, we believe that we have found significant
evidence of rotation in two out of three cores, while in the third object
detection of rotation might be hindered by projection effects (rotation axis
almost parallel to the line of sight). In any case, rotational motions
are undoubtly less prominent in these two cores than in those (G24.78+0.08
and G31.41+0.31) studied by Beltr\'an et al. (2004, 2005). Given the low
number statistics it is impossible to draw any conclusion out of this result.
However, one cannot rule out the possibility that this is an observational
effect. The distance to \gtt\ is 1.4 times larger than that of G24.78 and
G31.41, while the angular resolution of our observations is at least 2 times
worse than that of the PdBI observations by Beltr\'an et al. (2004, 2005).
Therefore, not only our observations are less sensitive to velocity gradients,
but this makes it also more difficult to disentangle the contribution by distinct
toroid-outflow systems overlapping in the same region -- especially when multiple
sources are present as in the case of \gte.
In conclusion, we believe that better angular and spectral resolutions
are needed to make any search for rotating disks/toroids successful.\par

Notwithstanding these caveats, it is worth stressing that
the mere existence of rotation in two of the HMCs studied by us, albeit
insufficient to guarantee support against gravitational forces, is an
important clue for the process of high-mass star formation.  
In addition, one cannot rule out the possibility that, 
on smaller scales than those imaged by us,
conservation of angular momentum might speed up the rotation,
and thus attain centrifugal equilibrium in a circumstellar disk.
This condition will be attained for radii satisfying the relation
\begin{equation}
R_{\rm eq} = R_{\rm eff} \, \left(\frac{\Mdyn}{\Mket}\right)^\frac{1}{4-p}
\end{equation}
whose derivation is given in Appendix~\ref{acenr}.
A minimum value of $R_{\rm eq}$ can be
obtained for $p=2.5$ (see e.g. Fontani et al. 2002), which gives 
$R_{\rm eq}/R_{\rm eff}\simeq 6\times10^{-3}$ and $5\times10^{-3}$
for \gte\ and \gtt, respectively.
Regions like these need sub-arcsec angular resolution and a higher density
tracer than \mcn\ to be investigated.

\section{Conclusions}

Using the OVRO, Nobeyama, and IRAM-PdB mm-interferometers,
we carried out intensive search for rotating toroids towards the massive YSOs in 
\gs, \gte, and \gtt\ which exhibit no or faint free-free emission (CTC97).
Our observations revealed that these objects are embedded in 
HMCs with masses of 95--380 \Msun\ and temperatures of 93--130 K, 
making the cores typical site of high-mass (proto)star formation.
All the 3 objects harbored in the HMCs are driving powerful (\Fco $\simeq$ $10^{-3}-10^{-2}$ 
\Msun \kms\ yr$^{-1}$) CO outflows.
However, the nature of the outflows in \gte\ and \gtt\ is unclear;
the origin of high velocity wing emission may be attributed to either
single or double outflow(s).
Such ambiguity made the interpretation of velocity gradients, 
identified through \mcn\ $K$-ladder line analysis, 
existing in the innermost densest part of the \gte\ and \gtt\ HMCs fairly difficult.
The velocity gradients are almost perpendicular to their molecular outflow axes, 
suggesting the presence of rotating, flattened structures.  
However, the corresponding dynamical masses are an order of magnitude smaller than 
the masses derived from 3\,mm dust continuum emission, 
which indicates turbulent pressure as the dominant support of the HMCs.
No conclusion could be reached for the third
source, \gs, as the putative rotation axis appears to lie close to the 
line-of-sight, thus making the detection of the rotation velocity very difficult
for projection effects.
Further higher resolution imaging will
allow us to establish the presence of rotation on a more solid ground.


\acknowledgments

The authors gratefully acknowledge the staff of OVRO, NRO, and IRAM
observatories, the Spitzer Science Center, and the GILDAS software group at IRAM.  
This publication used archival data from the {\it Spitzer Space Telescope} and {\it MSX}. 
The authors thank L. Testi, R. Bachiller, and M. Tafalla for their
early contribution to this study.
R. S. F. acknowledges A. I. Sargent, J. M.
Carpenter, H. Karoji, M. Hayashi, and S. S.  Hayashi for their continuous
encouragement and support, T. Y. Brooke for his generous help with data
reduction of the GLIMPSE images. 
During this work, M. T. B. was supported by MEC grant AYA2005-08523-C0, and
R.S.F. was supported by 
OVRO mm-array postdoctoral fellowship program under NSF grant AST 02-28955.

\appendix

\section{Centrifugal Radius}
\label{acenr}

Here we calculate an expression of the radius at which centrifugal equilibrium
can be attained inside a rotating, spherically symmetric core. The hypothesis
is that the rotation velocity \vrot\ observed at radius \Reff\
is not sufficient to sustain the core against gravitational forces thus
allowing contraction and speed up of the gas in the core until centrifugal
equilibrium is attained. If this occurs at radius \Req\ with velocity \veq,
conservation of angular momentum per unit mass gives the following
relation:
\begin{equation}
 \veq \Req = \vrot \Reff         \label{ecam}
\end{equation}
while the equilibrium condition turns into
\begin{equation}
 \veq^2 = \frac{G \, M(\Req)}{\Req}.         \label{ege}
\end{equation}
The mass $M(R)$ contained inside radius $R$ depends on the dependence of
the gas volume density $n$ on the radius. 
Assuming that $n$ varies with $R^{-p}$, one obtaines $M(R)\propto R^{3-p}$. 
Therefore, we can write the following relation
\begin{equation}
 M(\Req) = M(\Reff)\left(\frac{\Req}{\Reff}\right)^{3-p}   \label{emass}
\end{equation}
where $M(\Reff)=\Mdk$ measured by us from the mm continuum emission.

Substituting \veq\ from Eq.~(\ref{ecam}) and $M(\Req)$ from Eq.~(\ref{emass})
into Eq.~(\ref{ege}), and replacing \vrot\ with \Mdyn\ from the definition,
we find
\begin{equation}
\Mdyn = \frac{\Reff\,\vrot^2}{G},
\end{equation}
one consequently obtains the expression of \Req\ as a function of
measurable quantities:
\begin{equation}
\Req = \Reff \left(\frac{\Mdyn}{\Mdk}\right)^{\frac{1}{4-p}}.
\end{equation}



\clearpage
\begin{deluxetable}{ccccccc}
\tablewidth{0pt}
\tabletypesize{\scriptsize}
\tablecaption{Summary of Interferometric Continuum Emission Observations
\label{tbl:obs_cont}
}
\tablewidth{0pt}
\tablehead{
\colhead{\lw{Source}} &
\colhead{\lw{Telescope}} &  
\colhead{\lw{Frequency}} & 
\colhead{\lw{Bandwidth\tablenotemark{a}}} &   
\multicolumn{2}{c}{Synthesized Beam} &
\colhead{Image} \\
\cline{5-6}
\colhead{} &
\colhead{} & 
\colhead{} &
\colhead{} & 
\colhead{$\theta_{\rm maj}\times\theta_{\rm min}$} &
\colhead{P.A.} &
\colhead{RMS} \\
\colhead{} &
\colhead{} &
\colhead{(GHz)} & 
\colhead{(GHz)} & 
\colhead{(arcsec)} &
\colhead{(deg)} &
\colhead{(mJy \pbeam\ )} \\
}
\startdata
G\,16.59$-$0.05 & OVRO &  92.0\tablenotemark{b} & 3.5    & $2.31\times 1.75$ &  $-24$ & 0.50 \\
                & IRAM & 115.3                  & 0.32   & $6.31\times 4.10$ & $-167$ & 2.49 \\
G\,28.87$+$0.07 & OVRO &  92.0\tablenotemark{b} & 3.5    & $2.10\times 1.80$ &  $-74$ & 0.51 \\
                & IRAM & 115.3                  & 0.32   & $4.97\times 4.15$ &  $+23$ & 1.59 \\
G\,23.01$-$0.41 &  NMA &  98.8\tablenotemark{c} & 1.0    & $2.69\times 1.93$ &  $-14$ & 0.99 \\
                &  NMA & 110.0\tablenotemark{c} & 1.0    & $2.40\times 1.69$ &  $-12$ & 1.35 \\
                & IRAM & 115.3                  & 3.5    & $5.50\times 4.17$ & $-164$ & 1.90 \\
\enddata
\tablenotetext{a}{Effective bandwidth of line-free channels for obtaining continuum image.}
\tablenotetext{b}{Center frequency of both the side bands.}
\tablenotetext{c}{Center frequency of each single side band.}
\end{deluxetable}

\clearpage
\begin{deluxetable}{crrccccc}
\tablewidth{0pt}
\tabletypesize{\scriptsize}
\tablecaption{Summary of Interferometric Molecular Line Observations
\label{tbl:obs_lin}
}
\tablewidth{0pt}
\tablehead{
\colhead{\lw{Source}} &
\colhead{\lw{Line}} &
\colhead{Tracking} & 
\colhead{\lw{Telescope}} &   
\colhead{Velocity} &
\multicolumn{2}{c}{Synthesized Beam} &
\colhead{Image\tablenotemark{a}} \\
\cline{6-7}
\colhead{} &
\colhead{} &
\colhead{Frequency} & 
\colhead{} &
\colhead{Resolution} & 
\colhead{$\theta_{\rm maj}\times\theta_{\rm min}$} &
\colhead{P.A.} &
\colhead{RMS} \\
\colhead{} &
\colhead{} & 
\colhead{(MHz)} &
\colhead{} & 
\colhead{(\kms\ )} & 
\colhead{(arcsec)} &
\colhead{(deg)} &
\colhead{(mJy \pbeam\ )} \\
}
\startdata
G\,16.59$-$0.05 & \HCO\ (1--0)    &  89188.518 & OVRO & 1.68 & $3.26\times 1.95$ &  $-10$ & 51 \\
         & \mcn\ (5--4)           &  90584.244 & OVRO & 1.63 & $2.46\times 1.96$ &  $-27$ & 18 \\
         & \NtwoH\ (1--0)         &  93173.900 & OVRO & 0.80 & $2.45\times 1.85$ &  $-46$ & 27 \\
         & \co\ (1--0)            & 115271.204 & IRAM & 1.63 & $6.31\times 4.10$ & $-167$ & 43 \\
G\,28.87$+$0.07 & \HCO\ (1--0)    &  89188.518 & OVRO & 1.68 & $3.51\times 2.33$ &   $-9$ & 23 \\
         & \mcn\ (5--4)           &  90584.244 & OVRO & 1.63 & $2.47\times 2.05$ &  $-61$ & 18 \\
         & \NtwoH\ (1--0)         &  93173.900 & OVRO & 0.80 & $2.33\times 1.58$ &  $-71$ & 28 \\
         & \co\ (1--0)            & 115271.204 & IRAM & 1.63 & $4.58\times 4.58$ &  $-21$ & 24 \\
G\,23.01$-$0.41 & \CeO\ (1--0)    & 109782.156 & NMA  & 5.44 & $7.44\times 5.52$\tablenotemark{c} &  $-21$ & 25 \\
         & HNCO $5_{05}-4_{04}$   & 109905.753 & NMA  & 5.44 & $7.44\times 5.52$\tablenotemark{c} &  $-21$ & 27 \\
         & \mcn\ (6--5)           & 109997.353 & NMA  & 0.34\tablenotemark{b} & $2.86\times 1.67$ &   $-9$ & 43 \\
         & \tCO\ (1--0)           & 110201.352 & NMA  & 5.44 & $7.44\times 5.52$\tablenotemark{c} &  $-21$ & 32 \\
         & \co\ (1--0)            & 115271.204 & IRAM & 0.41 & $5.50\times 4.17$ & $-165$ & 43 \\
\enddata
\tablenotetext{a}{Typical Image noise level with the velocity resolution shown in the column 5.}
\tablenotetext{b}{Effective resolution after smoothing.}
\tablenotetext{c}{Received with the same intermediate-frequency, leading to the identical
beam size for the lines.}
\end{deluxetable}

\clearpage
\begin{deluxetable}{lcccccrrr}
\tablewidth{0pt}
\tabletypesize{\scriptsize}
\tablecaption{Properties of 3\,mm Continuum Emission
\label{tbl:res_cont}
}
\tablewidth{0pt}
\tablehead{
\colhead{\lw{Source}} &
\colhead{$d$\tablenotemark{a}} &
\colhead{$\nu$} &
\colhead{} &
\multicolumn{2}{c}{Peak Position} &
\colhead{} &
\colhead{\Reff\tablenotemark{b}} &
\colhead{$S_\nu$\tablenotemark{c}} \\
\cline{5-6}
\cline{6-7}
\colhead{} &
\colhead{(kpc)} &
\colhead{(GHz)} & 
\colhead{} &
\colhead{R.A.(J2000)} &
\colhead{Dec.(J2000)} &
\colhead{} &
\colhead{(pc)} &
\colhead{(mJy)} \\
}
\startdata
\gs\   &  4.7 &  92.0 & & 18:21:09.06 & $-$14:31:47.9 & & 0.033 & 37.5$\pm$7.5 \\
       &      & 115.3 & & 18:21:09.06 & $-$14:31:47.5 & & 0.055 & 79.6$\pm$16  \\
\hline
\gte\  &  7.4 &  92.0 & & 18:43:46.25 & $-$03:35:29.9 & & 0.029 & 11.3$\pm$2.3 \\
       &      & 115.3 & & 18:43:46.31 & $-$03:35:29.2 & & 0.102 & 42.6$\pm$8 \\
\gte B &      &  92.0 & & 18:43:46.19 & $-$03:35:12.9 & & 0.033 &  1.7$\pm$0.3 \\
\gte C &      &  92.0 & & 18:43:46.31 & $-$03:35:15.4 & & $\sim 0.01$ &  1.5$\pm$0.4 \\
\gte D &      & 115.3 & & 18:43:47.09 & $-$03:35:32.0 & & 0.030\tablenotemark{d}   & 0.80$\pm$0.2 \\
\hline
\gtt\  & 10.7 &  98.8 & & 18:34:40.29 & $-$09:00:38.1 & & 0.068 & 27.8$\pm$5.6 \\
       &      & 110.0 & & 18:34:40.29 & $-$09:00:38.6 & & 0.055 & 31.6$\pm$6.3 \\
       &      & 115.3 & & 18:34:40.30 & $-$09:00:38.2 & & 0.103 & 33.9$\pm$6.8 \\
\enddata
\tablecomments{See Table \ref{tbl:obs_cont} for the parameters of observations and $\S$\ref{ss:contmaps}
for the adopted detection criteria.}
\tablenotetext{a}{Distance to the region, see $\S$\,2 of CTC97.}
\tablenotetext{b}{Effective radius, see $\S$\ref{ss:contmaps} for the definition.}
\tablenotetext{c}{Total flux density integrated over the region enclosed by the 5\sgm\ level contour.}
\tablenotetext{d}{$\sqrt{\cal{A}/\pi}$ for the area encompassed by the 
5\sgm\ contour is given because the area is smaller than the beam size (Table \ref{tbl:obs_cont}).}
\end{deluxetable}

\clearpage
\thispagestyle{empty}
\voffset=20 mm
\begin{deluxetable}{ccccccccccccccccccccccccc}
\tablewidth{0pt}
\tabletypesize{\scriptsize}
\rotate
\tablecaption{Flux Densities of the 3 HMCs at Infrared Wavelengths
\label{tbl:IRfluxes}
}
\tablewidth{0pt}
\tablehead{
\colhead{} &
\colhead{} &
\multicolumn{2}{c}{21.3\,\um} &
\colhead{} &
\multicolumn{2}{c}{14.7\,\um} &
\colhead{} &
\multicolumn{2}{c}{12.1\,\um} &
\colhead{} &
\multicolumn{2}{c}{8.3\,\um} &
\colhead{} &
\multicolumn{2}{c}{8.0\,\um} &
\colhead{} &
\multicolumn{2}{c}{5.8\,\um} &
\colhead{} &
\multicolumn{2}{c}{4.5\,\um} &
\colhead{} &
\multicolumn{2}{c}{3.6\,\um} \\
\cline{3-4}
\cline{6-7}
\cline{9-10}
\cline{12-13}
\cline{15-16}
\cline{18-19}
\cline{21-22}
\cline{24-25}
\colhead{Source} &
\colhead{} & 
\colhead{$S$} &
\colhead{\lw{Note}} & 
\colhead{} &
\colhead{$S$} &
\colhead{\lw{Note}} & 
\colhead{} &
\colhead{$S$} & 
\colhead{\lw{Note}} &
\colhead{} &
\colhead{$S$} & 
\colhead{\lw{Note}} &
\colhead{} & 
\colhead{$S$} &
\colhead{\lw{Note}} &
\colhead{} & 
\colhead{$S$} &
\colhead{\lw{Note}} & 
\colhead{} &
\colhead{$S$} &
\colhead{\lw{Note}} & 
\colhead{} &
\colhead{$S$} & 
\colhead{\lw{Note}} \\
\colhead{} &
\colhead{} & 
\colhead{(Jy)} &
\colhead{} & 
\colhead{} &
\colhead{(Jy)} &
\colhead{} & 
\colhead{} &
\colhead{(Jy)} & 
\colhead{} &
\colhead{} &
\colhead{(Jy)} & 
\colhead{} &
\colhead{} & 
\colhead{(Jy)} &
\colhead{} &
\colhead{} & 
\colhead{(Jy)} &
\colhead{} & 
\colhead{} &
\colhead{(Jy)} &
\colhead{} & 
\colhead{} &
\colhead{(Jy)} & 
\colhead{} \\
}
\startdata
G\,16.59$-$0.05  & &  7.7 & 1,2 & &     1.8 & 1,2 & &     4.1 & 1,2,3 & &     2.4 & 1,2 & & 6.8E--2 & 2,4 & & 3.3E--2 & 2,4 & & 1.4E--2 & 2,4 & & \nodata & 5 \\
G\,28.87$+$0.07  & & 14.7 & 1,2 & &     6.8 & 1,2 & &     5.1 & 1,2   & &     2.8 & 1,2 & & 1.3     & 2,4 & & 8.0E--1 & 2,4 & & 4.0E--1 & 2,4 & & 6.0E--2 & 2,4 \\
G\,23.01$-$0.41  & &  4.4 & 1,2 & & $<0.51$ & 1,6 & & $<0.86$ & 1,6   & & $<0.20$ & 1,6 & & 1.1E--1 & 2,4 & & 1.1E--1 & 2,4 & & 6.3E--2 & 2,4 & & 1.0E--2 & 2,4 \\
\enddata
\tablecomments{
1: MSX data,
2: integrated inside the 50\% level contour of the data, 
3: contaminated with the nearby source(s),
4: GLIMPSE data,
5: impossible to define a point source (see Figure \ref{fig:contmaps}, and
6: $3\sigma$ upper limit.
See $\S$\ref{ss:sed} for the details.}
\end{deluxetable}

\clearpage
\voffset=0 mm
\begin{deluxetable}{ccc}
\tablewidth{0pt}
\tabletypesize{\scriptsize}
\tablecaption{Core Mass Estimated from 3\,mm Continuum Emission
\label{tbl:Mcore}
}
\tablewidth{0pt}
\tablehead{
\colhead{\lw{Source}} &
\colhead{$M_{\rm dust}$\tablenotemark{a}} &
\colhead{$\langle N(\rm H_2)\rangle$\tablenotemark{b}} \\
\colhead{} &
\colhead{(\Msun)} &
\colhead{($\times 10^{23}$ \cmq)} \\
}
\startdata
\gs\  &  95$\pm$6  & 3.0$\pm$0.9 \\
\gte\ & 100$\pm$7  & 4.2$\pm$1.3 \\
\gtt\ & 380$\pm$22 & 3.6$\pm$1.1 \\
\enddata
\tablecomments{See $\S$\ref{ss:sed} for the details.}
\tablenotetext{a}{Mass of the core estimated from the 3\,mm continuum flux density
shown in Table \ref{tbl:res_cont}. 
We assumed that \Tdust\ is equal to \Trot \ of \mcn\ emission 
(Table \ref{tbl:res_mcn}).}
\tablenotetext{b}{Mean H$_2$ column density calculated over the area
encompassing the emission inside the 5\sgm\ contour.}
\end{deluxetable}

\begin{deluxetable}{ccccc}
\tablewidth{0pt}
\tabletypesize{\scriptsize}
\tablecaption{Properties of the Large-Scale Clumps Surrounding the 3 HMCs
\label{tbl:clumps}
}
\tablewidth{0pt}
\tablehead{
\colhead{\lw{Source}} &
\colhead{} &
\colhead{\lw{Probe}} &
\colhead{$R_{\rm eff}$\tablenotemark{a}} &
\colhead{\Mlte\tablenotemark{b}} \\
\colhead{} &
\colhead{} &
\colhead{} &
\colhead{(pc)} &
\colhead{(\Msun)} \\
}
\startdata
\gs\  & & \HCO\ (1--0)   & 0.070 &       440 -- 1500 \\ 
      & & \NtwoH\ (1--0) & 0.15  &               540 \\ 
\gte\ & & \HCO\ (1--0)   & 0.12  &        84 --  360 \\ 
\gtt\ & & \tCO\ (1--0)   & 0.426 &    $1\times 10^4$ \\ 
      & & \CeO\ (1--0)   & 0.174 &    $2\times 10^4$ \\ 
\enddata
\tablecomments{For details, see $\S$\ref{ss:totmaps}.}
\tablenotetext{a}{Effective radius, see $\S$\ref{ss:contmaps} for the definition.}
\tablenotetext{b}{\,LTE-mass obtained by integrated the emission over
the area enclosed by the 5\sgm\ level contours.}
\end{deluxetable}

\clearpage
\begin{deluxetable}{cccccc}
\tablewidth{0pt}
\tabletypesize{\scriptsize}
\tablecaption{Parameters of Molecular Outflows
\label{tbl:outflow}
}
\tablewidth{0pt}
\tablehead{
\colhead{Source} &
\colhead{Lobe} &
\colhead{$M_{\rm lobe}$\tablenotemark{a}} &
\colhead{$t_{\rm d}$\tablenotemark{b}} &
\colhead{$\dot{M}_{\rm flow}$\tablenotemark{c}} &
\colhead{$F_{\rm CO}$\tablenotemark{d}} \\
\colhead{} &
\colhead{} &
\colhead{(\Msun)} & 
\colhead{(yrs)} & 
\colhead{(\Msun\ yr$^{-1}$)} &
\colhead{(\Msun\ \kms\ yr$^{-1}$)} \\
}
\startdata
G\,16.59$-$0.05 & Blue    &  21$\pm$11 & 2.7$\times 10^4$ & (2$\pm$1)$\times 10^{-4}$ & (9$\pm$5)$\times 10^{-3}$ \\
                & Red     &   7$\pm$4  & 1.4$\times 10^4$ & (5$\pm$3)$\times 10^{-4}$ & (9$\pm$5)$\times 10^{-3}$ \\
G\,28.87$+$0.07 & Blue    &  25$\pm$12 & 2.6$\times 10^4$ & (3$\pm$2)$\times 10^{-4}$ & (2$\pm$1)$\times 10^{-2}$ \\
                & Red     &  90$\pm$40 & 3.1$\times 10^4$ & (2$\pm$1)$\times 10^{-3}$ & (5$\pm$3)$\times 10^{-2}$ \\
G\,23.01$-$0.41 & Blue SW &  17$\pm$12 & 2.5$\times 10^4$ & (9$\pm$7)$\times 10^{-4}$ & (2.5$\pm$1.5)$\times 10^{-2}$ \\
                & Blue NE &  14$\pm$10 & 2.8$\times 10^4$ & (6$\pm$3)$\times 10^{-4}$ & (1.9$\pm$1.2)$\times 10^{-2}$ \\
                & Red SW  & 140$\pm$70 & 3.4$\times 10^4$ & (4$\pm$3)$\times 10^{-3}$ & (1.3$\pm$0.7)$\times 10^{-1}$ \\
                & Red NE  &   4$\pm$3  & 3.4$\times 10^4$ & (1.2$\pm$0.7)$\times 10^{-4}$ & (3.5$\pm$3.0)$\times 10^{-3}$ \\
\enddata
\tablecomments{Outflow characteristic estimated from the \co\ (1--0) data, 
see $\S$\ref{ss:outflow}. All the errors are considered only for the \Tex\ uncertainty.}
\tablenotetext{a}{Outflow lobe mass.}
\tablenotetext{b}{Dynamical time scale.}
\tablenotetext{c}{Outflow mass loss rate.}
\tablenotetext{d}{Outflow momentum flux.}
\end{deluxetable}

\begin{deluxetable}{ccccccccc}
\tablewidth{0pt}
\tabletypesize{\scriptsize}
\tablecaption{Results of \mcn\ Spectra Analysis
\label{tbl:mcnsp}
}
\tablewidth{0pt}
\tablehead{
\colhead{\lw{Source}} &
\colhead{\lw{Transition}} &
\colhead{$V_{\rm LSR}$\tablenotemark{a}} &
\colhead{$\Delta v$\tablenotemark{b}} & 
\colhead{} &
\multicolumn{4}{c}{$\int T_{\rm sb}dv$ (K$\cdot$\kms)} \\
\cline{5-9}
\colhead{} &
\colhead{} &
\colhead{(\kms)} & 
\colhead{(\kms)} &
\colhead{} &
\colhead{$K = 0$} &
\colhead{1} &
\colhead{2} &
\colhead{3} \\
}
\startdata
G\,16.59$-$0.05 & $J=$\,5--4 &  59.9$\pm$0.2 & 5.6$\pm$0.2 & & 20.5$\pm$1.2 & 14.1$\pm$1.2 & 10.5$\pm$1.1 & 13.7$\pm$1.1 \\
G\,28.87$+$0.07 & $J=$\,5--4 & 103.5$\pm$0.3 & 9.1$\pm$0.4 & & 12.9$\pm$1.4 & 11.5$\pm$1.5 &  7.8$\pm$0.9 &  7.7$\pm$0.9 \\
G\,23.01$-$0.41 & $J=$\,6--5 &  77.4$\pm$0.1 & 8.3$\pm$0.2 & & 16.2$\pm$0.8 & 16.7$\pm$0.8 & 15.0$\pm$0.6 & 13.8$\pm$0.6 \\
\enddata
\tablecomments{For details, see $\S\ref{ss:rd}$.}
\tablenotetext{a}{Centroid velocity from the multiple-Gaussian profile fitting. 
We adopted the \Vlsr\ in column 3 as the \Vsys\ of the HMCs.}
\tablenotetext{b}{\,Line width in FWHM.}
\end{deluxetable}

\begin{deluxetable}{cccccccccc}
\tablewidth{0pt}
\tabletypesize{\scriptsize}
\tablecaption{Summary of Methyl Cyanide RD Analysis
\label{tbl:res_mcn}
}
\tablewidth{0pt}
\tablehead{
\colhead{} &
\colhead{$R_{\rm eff}$\tablenotemark{a}} &
\colhead{} & 
\multicolumn{2}{c}{Temperature\tablenotemark{b}(K)} &
\colhead{} & 
\multicolumn{1}{c}{$N_{\rm CH_3CN}$\tablenotemark{c}} &
\colhead{} &
\colhead{$X_{\rm \mcn}$\tablenotemark{d}} \\
\cline{4-5}
\colhead{Source} &
\colhead{\lw{(pc)}} & 
\colhead{} &
\colhead{\lw{\Trot (\mcn )}} & 
\colhead{\lw{\Tk (\amm )}} & 
\colhead{} & 
\colhead{\lw{(\cmq )}} &
\colhead{} &
\colhead{\lw{($\times 10^{-10}$)}} \\
}
\startdata
\gs\  & 0.029 & & 130$^{+36}_{-23}$ & 54 & & (6.9$^{+2.9}_{-1.5}$)$\times 10^{14}$ & & 20$\pm$7 \\
\gte\ & 0.042 & &  93$^{+31}_{-18}$ & 37 & & (3.1$^{+1.3}_{-0.6}$)$\times 10^{14}$ & &  7$\pm$2 \\
\gtt\ & 0.065 & & 121$^{+17}_{-13}$ & 58 & & (4.6$^{+0.9}_{-0.5}$)$\times 10^{14}$ & & 13$\pm$4 \\
\enddata
\tablecomments{For details, see $\S\S$\ref{ss:rd} and \ref{ss:Xmcn}.}
\tablenotetext{a}{Effective radius for the \mcn\ $K=0+1$ emitting region,
see $\S$\ref{ss:contmaps} for the definition.}
\tablenotetext{b}{Rotational temperature for \mcn\ ($\S$\ref{ss:rd}) 
and kinetic temperature for \amm\ (CTC97).}
\tablenotetext{c}{Total column density of \mcn.}
\tablenotetext{d}{Fractional abundance of \mcn\ obtained from comparisons
with the mean \meanNmol\ in Table \ref{tbl:Mcore}.}
\end{deluxetable}

\begin{deluxetable}{ccccccc}
\tablewidth{0pt}
\tabletypesize{\scriptsize}
\tablecaption{Properties of the Central Regions of the 3 HMCs
\label{tbl:CoreCenter}
}
\tablewidth{0pt}
\tablehead{
\colhead{Source} &
\colhead{\Mdust\tablenotemark{a}} &
\colhead{\dVint\tablenotemark{b}} & 
\colhead{\Mvir\tablenotemark{c}} &
\colhead{\Vrot\tablenotemark{d}} & 
\colhead{\Mdyn\tablenotemark{e}} & 
\colhead{\Mket\tablenotemark{,f}} \\
\colhead{} &
\colhead{(\Msun)} &
\colhead{(\kms)} & 
\colhead{(\Msun)} & 
\colhead{(\kms)} & 
\colhead{(\Msun)} & 
\colhead{(\Msun)} \\
}
\startdata
\gs\  &  95$\pm$6  & 5.4$\pm$0.3 &  180$\pm$15 &     \nodata\ & \nodata\ & \nodata\ \\
\gte\ & 100$\pm$7  & 9.1$\pm$0.4 &  780$\pm$78 & $\simeq 0.5$ &  3$\pm$2 &  55$\pm$3 \\
\gtt\ & 380$\pm$22 & 8.6$\pm$0.2 & 1100$\pm$40 & $\simeq 0.6$ &  8$\pm$3 & 340$\pm$20 \\
\enddata
\tablecomments{For details, see $\S$\ref{ss:VelSt}.}
\tablenotetext{a}{From Table \ref{tbl:Mcore}.}
\tablenotetext{b}{Intrinsic velocity width (\dVint) in FWHM after deconvolving the instrumental
velocity resolution (Table \ref{tbl:obs_lin}) where \dVint\ is measured over the region enclosed
by the 50\% level contour of the emission. Here, contributions of thermal gas
motions are 0.38, 0.32, and 0.37 \kms\ for \gs, \gte, and \gtt, respectively, 
when we assume that kinematical temperature (\Tk) of the gas is equal to the
mean \Trot\ in Table \ref{tbl:res_mcn}.}
\tablenotetext{c}{Virial mass over the 50\% area of \mcn\ $K=0+1$ emission.
We adopted \Reff\ in Table \ref{tbl:res_mcn} for calculating \Mvir.}
\tablenotetext{d}{Rotation velocity assuming that the velocity gradient is
produced by rotation of the core.}
\tablenotetext{e}{Dynamical mass.}
\tablenotetext{f}{Core mass calculated from the 3\,mm continuum flux density
over the region enclosed by the 5$\sigma$ level contour of the \mcn\ $K=3$ emission.}
\end{deluxetable}



\clearpage

\clearpage

\clearpage

\clearpage

\clearpage

\clearpage

\clearpage

\clearpage

\clearpage

\end{document}